\documentclass[11pt]{article}
\usepackage{comment,soul}
\usepackage[hidelinks]{hyperref}
\usepackage{enumitem}
\usepackage{amsmath,amssymb,epsf,cite,graphicx,subfigure}
\usepackage{amsmath,braket,tikzsymbols}
\usepackage[bbgreekl]{mathbbol}
\usepackage{comment}
\numberwithin{equation}{section}

\definecolor{airforceblue}{rgb}{0.36, 0.54, 0.66}
\newcommand{\beq}{\begin{equation}}
\newcommand{\eeq}{\end{equation}}

\newcommand{\pd}{\partial}

\usepackage{todonotes}

\usepackage{simplewick}

\begin{document}
\baselineskip=15.5pt
\pagestyle{plain}
\setcounter{page}{1}

\begin{center}

{\LARGE \bf Quantizing Carrollian field theories}
\vskip 1cm

\textbf{Jordan Cotler$^{1,a}$, Kristan Jensen$^{2,b}$, Stefan Prohazka$^{3,c}$, \\ Amir Raz$^{4,d}$, Max Riegler$^{5,e}$, and Jakob Salzer$^{6,f}$}

\vspace{0.5cm}

{\it ${}^1$ Department of Physics, Harvard University, Cambridge, MA 02138, USA \vspace{.3cm}}

{\it${}^2$Department of Physics and Astronomy, University of Victoria, Victoria, BC V8W 3P6, Canada\vspace{.3cm}}

{\it ${}^3$University of Vienna, Faculty of Physics, Mathematical Physics, Boltzmanngasse 5, 1090, Vienna, Austria\vspace{.3cm}}

{\it ${}^4$Theory Group, Department of Physics, University of Texas, Austin, TX 78712, USA\vspace{.3cm}}

{\it ${}^5$Quantum Technology Laboratories GmbH, Clemens-Holzmeister-Straße 6/6, 1100 Vienna, Austria\vspace{.3cm}}

{\it ${}^6$Universit\'e Libre de Bruxelles and International Solvay Institutes, ULB-Campus Plaine CP231, B-1050 Brussels, Belgium}

\vspace{0.3cm}

{\tt ${}^a$jcotler@fas.harvard.edu, \small ${}^b$kristanj@uvic.ca, ${}^c$stefan.prohazka@univie.ac.at, ${}^d$araz@utexas.edu, ${}^e$rieglerm@hep.itp.tuwien.ac.at, ${}^f$jakob.salzer@ulb.be\\}

\medskip

\end{center}

\vskip1cm

\begin{center}
{\bf Abstract}
\end{center}
\hspace{.3cm} 
Carrollian field theories have recently emerged as a candidate dual to flat space quantum gravity. We carefully quantize simple two-derivative Carrollian theories, revealing a strong sensitivity to the ultraviolet. They can be regulated upon being placed on a spatial lattice and working at finite inverse temperature. Unlike in conventional field theories, the details of the lattice-regulated Carrollian theories remain important at long distances even in the limit that the lattice spacing is sent to zero. We use that limit to define interacting continuum models with a tractable perturbative expansion. The ensuing theories are those of generalized free fields, with non-Gaussian correlations suppressed by positive powers of the lattice spacing, and an unbroken supertranslation symmetry. 
\newpage

\tableofcontents

\section{Introduction}

What are the holographic duals of consistent theories of flat space quantum gravity? There have been recent suggestions~\cite{Donnay:2022aba,Bagchi:2022emh,Donnay:2022wvx,Have:2024dff} (see~\cite{Donnay:2023mrd} for a review) that the duals are conformal Carrollian field theories, building on the observation that the Poincar\'e group acts as the conformal version of the Carroll group on null infinity. Despite being somewhat exotic, Carrollian theories date back to the 1960's as the ultrarelativistic limit (i.e.\ $c\to 0$) of relativistic field theories\cite{Levy1965,SenGupta1966OnAA}. See also~\cite{Klauder:2000ud} and references therein. In that limit all excitations move at the speed of light and so are immobile from the point of view of a comoving observer.\footnote{The name is a reference to Lewis Carroll's \emph{Alice in Wonderland}: ``My dear, here we must run as fast as we can, just to stay in place.''} At this time there is no known example of Carrollian holography in string theory. Part of the difficulty in identifying a putative dual is that Carrollian theories have somewhat strange behavior. The purpose of this manuscript is to demystify some of this strange behavior by performing careful path integral treatments of these theories. As much as possible we attempt to obtain general lessons that apply to a range of Carrollian theories, with the hope of helping focus the ongoing search for flat space duals.\footnote{Our manuscript builds upon a number of recent works discussing the quantization of Carrollian field theories. For a partial list see~\cite{Bagchi:2016bcd,Chen:2021xkw,Bagchi:2022eui,Donnay:2022aba,Bagchi:2022emh,Donnay:2022wvx,Mehra:2023rmm,Banerjee:2023jpi,Figueroa-OFarrill:2023qty,Bekaert:2024itn,Chen:2024voz}, and especially~\cite{deBoer:2023fnj}.}

Our line of inquiry begins with two-point functions in Carrollian theories in $d$ spatial dimensions, which are constrained by Carroll symmetry to take the form
\beq
\label{E:twoPoint}
	\langle \mathcal{O}(u_1,x_1)\mathcal{O}(u_2,x_2)\rangle = F(u_1-u_2)\delta(x_1-x_2) + G(x_1-x_2)\,,
\eeq
where here and throughout $u$ is time. The functions $F$ and $G$ are further constrained in a conformally invariant theory. The ultralocal structure proportional to $F$ is sometimes called ``electric'' and the time-independent correlation $G$ is sometimes called ``magnetic.'' There are simple field theories that we can call ``purely electric'' and ``purely magnetic'' that each reproduce exactly one of these types of correlations. 

An example of an ``electric'' theory is a real scalar $\phi$ with action~$S_E = \int du \,d^dx \left( \frac{1}{2}(\partial_u\phi)^2 - \frac{m^2}{2}\phi^2\right)$.  An example of a ``magnetic'' theory has two fields $\chi$ and $Q$ and an action $S_M = \int du\,d^dx \left( \chi \partial_u Q - \mathcal{H}\right)$, where $\mathcal{H}$ is a function of $\chi$ and its spatial derivatives. The simplest examples have $\mathcal{H} \sim |\vec{\nabla}\chi|^2 + \chi^2$. These simple ``electric'' and ``magnetic'' theories do not just have a Carroll boost symmetry, namely invariance under $u \to u + \vec{a}\cdot \vec{x}$, but a more general invariance under ``supertranslations'' $u\to u + f(\vec{x})$, whose corresponding Ward identity implies the absence of energy flux. One expects such a supertranslation symmetry in a putative holographic field theory dual to flat space quantum gravity in three or four spacetime dimensions, since in those cases the bulk has an infinite-dimensional BMS symmetry.  One further expects that the supertranslation symmetry is spontaneously broken.

Both theories mentioned above have a well-defined canonical quantization with a spectrum of finite-energy eigenstates. Nevertheless the electric theory is sensitive to the ultraviolet, and the magnetic theory to the infrared. These sensitivities proliferate when coupling an electric theory to a magnetic one. The major goals of this work are to explain this sensitivity and to develop practical strategies to compute observables in spite of it.

The reason for these sensitivities is easy to explain. Electric theories like the one above, or more general ones with a potential $V(\phi)$, are quite simple: in a lattice approximation each site is equipped with a quantum mechanics, and the sites are decoupled from each other. This leads to a thermal free energy that goes as the number of sites, $\frac{V}{a^d}$, with $V$ the spatial volume and $a$ the lattice spacing. The continuum limit is $a\to 0$ with $V$ fixed, leading to a UV-sensitive density of states. This is a simple form of UV/IR mixing, reminiscent of the recent literature on fractons.\footnote{In fact there is an analogy between theories with a conserved dipole moment, considered in the fracton literature, and those with Carroll symmetry. The idea is to exchange the charge and dipole moment with the Carroll energy and center of mass respectively~\cite{Bidussi:2021nmp,Marsot:2022imf,Figueroa-OFarrill:2023vbj}. Such a mapping exists at the level of the algebra of symmetry charges but not for field theories, since doing so would entail trading internal symmetries for spacetime symmetries.  However, a study of the representation theory of these symmetries implies that massive Carroll particles are immobile for the same reason that isolated charges are immobile in theories with dipole conservation.} 

The simple electric theories above have additional UV sensitivity. We can already see this in the Gaussian model with $V = \frac{m^2}{2}\phi^2$. Despite the lattice-sensitive density of states, correlation functions of $\phi$ are perfectly finite, giving rise to the ultralocal structure $F$ in the parameterization~\eqref{E:twoPoint} of its two-point function. However, composite operators $\phi^n$ have divergent correlations for $n>1$, with e.g.\ $\langle \phi^n(u,x) \phi^n(0,0)\rangle \propto (\delta^d(0))^{n-1}$. In the lattice regularized theory we can use the lattice spacing to define normalized versions of composite operators, e.g.\ $a^{\frac{(n-1)d}{2}}\phi^n$, which have finite two-point functions in the $a\to 0$ limit. Thanks to the powers of $a$ appearing in the normalization, these normalized operators have exactly Gaussian correlations in the limit. The same result applies when the on-site quantum mechanics has a non-quadratic potential but can still be treated perturbatively. Demanding that the on-site theory remain perturbative, we uncover a scaling limit whereby interactions, e.g.\ monomials $\delta V \sim \phi^p$ with $p>2$, must also be correctly normalized, suppressed by an appropriate power of $a$. The resulting continuum theory is perturbative, has unbroken supertranslation symmetry, and only Gaussian correlations.

In fact the simple electric theories are invariant under more symmetry than the supertranslations mentioned above. They are also invariant under volume preserving diffeomorphisms (VPDs), that preserve the spatial volume density $d^dx$. This symmetry is undesirable from the point of view of flat space holography and we discuss how to break it by higher derivative operators or by coupling matter fields to electric gauge theory. The former appears to lead to the spontaneous breaking of Carroll symmetry, while the latter preserves it.

In addition to electric theories of matter we consider an ``electric'' version of abelian gauge theory, with an action $S_E' = \frac{1}{2}\int du \,d^dx\, E^2$ with $E_i = \partial_u A_i-\partial_i A_u$\,. This theory is surprisingly subtle. As a warmup we consider the theory of a compact Carrollian scalar $\phi$ with action $S = \frac{1}{2}\int du \,d^dx \,(\partial_u\phi)^2$ with $\phi \sim \phi + 2\pi R$. In a lattice regularization we have the quantum mechanics of a rotor at each lattice site with a moment of inertia $a^d R^2$, which leads to a spectrum and correlation functions that differ significantly from those predicted by a na\"{i}ve path integral treatment. For example, for $R$ finite as $a\to 0$ the scalar theory has a one-dimensional Hilbert space of low-energy states (i.e.\ with energies below the lattice scale $E \ll \frac{1}{a^d}$). The culprit for this discrepancy is the sum over discontinuous field configurations, as in the recent literature on fractons~\cite{Seiberg:2020bhn}, and in particular the sum over modes which wind around the thermal circle discontinuously as a function of position. These configurations are absent in the na\"{i}ve path integral treatment of this model but are accounted for in our lattice quantization. From here we go on to discuss Carrollian electromagnetism, which in a certain gauge is $d$ copies of this compact scalar theory, subject to the Gauss' Law constraint. That constraint allows for local fluctuations of the electric field in $d>1$ spatial dimensions, the Carrollian analogue of transverse polarizations. Later we find it convenient to study this theory in the scaling limit where the bare electromagnetic coupling is $a^{d/2} g$ with the rescaled coupling $g$ held finite. The ensuing continuum limit is gapless.

By studying these examples we develop a diagrammatic prescription for computing correlation functions in electric theories directly in the continuum limit. This prescription is valid when the model has saddle points that are insensitive to details of the lattice. So the compact scalar cannot be treated this way, but a non-compact scalar with a finite on-site potential can.

Meanwhile, magnetic theories seem to have no dynamics: in the $S_M$ example above, $Q$ acts as a Lagrange multiplier enforcing that $\chi$ depends on space alone, i.e.\ $\chi$ is locally conserved. As a result the two-point function of $\chi$ is time-independent, leading to the tensor structure $G$ in the parameterization~\eqref{E:twoPoint}. However, the theory has a non-trivial spectrum of energy eigenstates and so non-trivial and finite transition amplitudes in the eigenbasis of the field operator $\hat{\chi}(x)$. This spectrum is continuous, and so if one considers instead the thermal partition function one finds a divergent result proportional to a field space Dirac delta function $\delta[0]$. This is really an IR divergence and appears universally in any quantum mechanical system with a continuous spectrum of energy eigenstates. Stripping off this normalization leads to a finite partition sum and vacuum-to-vacuum amplitude as well as a sensible dictionary between the $d+1$-dimensional magnetic theory and a $d$-dimensional Euclidean field theory. 

There are several natural IR-finite objects to study in magnetic theories. One is the aforementioned $S$-matrix. We can also consider thermal correlation functions at inverse temperature $\beta$, especially in the limit $\beta \to \infty$. For a magnetic theory that limit is a semiclassical one, with $\beta$ playing the role of $1/\hbar$. Correlations of the $\chi$'s and composite operators built from them become Gaussian in the $\beta\to\infty$ limit. Meanwhile correlations of $Q$, the conjugate of $\chi$, are infrared-divergent, analogous to those of the position operator in momentum eigenstates of a free quantum mechanical particle, while correlations of $\partial_u Q$ are in general IR-finite. While $\chi$ and composite operators built from it have magnetic correlations, $\partial_u Q$ may have both electric and magnetic parts.

We regard these basic electric and magnetic theories as building blocks which we can then put together to build richer Carrollian field theories. We study two examples in detail, finding similar physics in both, although in each case there is no conformal invariance. The first theory, which we call a Carrollian scalar Yukawa theory, couples a massive electric matter field $\phi$ to a magnetic theory $(\chi,Q)$ through Yukawa interactions $\sim \chi \phi^2$. The second theory is the Carrollian analogue of scalar QED, coupling a massive electric complex scalar to Carrollian electromagnetism. In both cases the coupling between building blocks breaks the VPD invariance of the electric sector described by $\phi$ and thereby generates correlations at nonzero spatial separation.

In the scalar Yukawa theory the electric building block $\phi$ is sensitive to the ultraviolet while the magnetic ingredients $(\chi,Q)$ are sensitive to the infrared. These sensitivities persist in the coupled theory. We are able to successfully define a perturbative interacting theory in the following way. As in the pure electric theory it is natural to work on a spatial lattice with a suitable $a \to 0$ limit. The various ultraviolet divergences we encounter are regulated by normalizing the ``electric operator'' $\phi^2$ appearing in the Yukawa interaction with a factor of $a^{d/2}$ as in our discussion above for the pure electric theory. As in the pure magnetic theory we also turn on a large but finite inverse temperature $\beta$ and we then take the limit $\beta\to \infty$. In this limit the scalar Yukawa theory has tractable and non-trivial perturbative interactions, only Gaussian correlations, unbroken supertranslation symmetry, and behaves like quantum mechanics under renormalization group flow to one-loop order. In other words there is no loop correction to scaling.

We find very similar statements throughout for Carrollian scalar QED, which in particular has a perturbative regime when the electromagnetic coupling scales as $a^{d/2}$ as $a\to 0$. 

In each case we study the Hamiltonian formulation in detail, and are able to reproduce our findings directly from a suitable quantum mechanical perturbation theory. These formulations are rather different from each other -- the scalar Yukawa theory has a direct interaction between the electric and magnetic degrees of freedom, while the interaction of Carrollian scalar QED is only through the Gauss' Law constraint -- and yet many features are shared.

The main lessons from our analysis are as follows.
\begin{enumerate}
	\item Carrollian theories are often sensitive to the UV and the IR, and exhibit UV/IR mixing. 
	\item This sensitivity and mixing is often tractable. We define perturbative, interacting two-derivative theories by first regulating with a spatial lattice and working at finite temperature, scaling couplings with appropriate powers of the lattice spacing, and then taking a continuum and zero temperature limit.
	\item The theories so obtained have unbroken supertranslation symmetry, can break ultralocality, but possess only Gaussian correlations. Moreover to one-loop order and in massive phases they resemble quantum mechanics in that they have classical renormalization group flow.
\end{enumerate}

On the one hand our work charts a path forward on the field theory side of a desired holographic duality between flat space quantum gravity and a Carrollian CFT. In particular our methods can be used to study more sophisticated models like non-abelian gauge theories coupled to matter. It should also be possible to study conformal theories, whereas we mostly focus on massive ones. On the other hand our work sharpens our understanding of the challenges to finding such a duality. The most obvious problems to overcome are that a dual pair must have its supertranslation symmetry broken to the subgroup of Carrollian boosts, and furthermore exhibit non-trivial non-Gaussianity. Perhaps these challenges are related.

The remainder of this manuscript is organized as follows. In Sections~\ref{S:electric} and~\ref{S:magnetic} we discuss simple electric and magnetic theories respectively, mostly demystifying their UV- and IR-sensitivities. In between we study compact electric scalars and Carrollian electromagnetism in Section~\ref{S:EM}. We couple these building blocks together in Section~\ref{S:electromagnetic}, and we wrap up with a Discussion focused on flat space holography in Section~\ref{sec:towards-carr-hologr}.

\section{``Electric theories''}
\label{S:electric}

\subsection{Two derivative models}
\label{S:electric2}

The simplest so-called ``electric theories''~\cite{Henneaux:2021yzg,deBoer:2021jej} are like those we discussed in the Introduction where the action lacks any spatial derivatives. For a single real scalar field an action of this sort is $S = \int du\, d^dx \left( \frac{1}{2}(\partial_u\phi)^2 - V(\phi)\right)$. These models sometimes go by the name of ``ultralocal field theories''~\cite{Klauder:2000ud} and one can land on them by a suitable $c\to 0$ limit of relativistic theories~\cite{Duval:2014uoa,Bergshoeff:2014jla}. They are Carroll-invariant because $\partial_u$ is invariant under supertranslations $u\to u + f(x)$. It is possible to write down Carroll-invariant deformations to these theories with spatial derivatives, but the requisite operators are at least quartic in the fundamental fields as well as in derivatives. Thus at the two derivative level we are dealing with Lagrangians with up to two time derivatives.

These seemingly innocuous theories realize UV/IR mixing. First consider the scalar theory with $V = \frac{m^2}{2}\phi^2$. It has a simple canonical quantization with regularly spaced energy eigenvalues $E = n m$ for $n=0,1,2,...$ and ladder operators $a^{\dagger}_{\vec{k}}$ for $\vec{k}$ any spatial momentum. 
In particular, $a^{\dagger}_{\vec{k}_1}\cdots a^{\dagger}_{\vec{k}_n}|0\rangle$ (with $|0\rangle$ the ground state) is an energy eigenstate of energy $n m$ for all choices of momenta $\{\vec{k}_1,\hdots,\vec{k}_n\}$, and so we have a model of ``flat bands'' since $a_{\vec{k}}^\dagger |0\rangle$ has the same energy for any momentum $\vec{k}$. While the spectrum is regular and finite, the density of states diverges, and so do simple ``global'' quantities like the thermal free energy $F$.  Explicitly,
\beq
	\frac{F}{VT} =\frac{1}{2} \sum_{n\in \mathbb{Z}}\int \frac{d^dk}{(2\pi)^d} \log (\omega_n^2 + m^2)\,, \qquad \omega_n = \frac{2\pi n}{\beta}\,,
\eeq
with $V$ the volume and $T$ the temperature, and $F$ diverges on account of the integral over spatial momenta. One might also worry about the vacuum physics of the model with perturbatively small polynomial interactions: the tree-level propagator and vertices are momentum-independent, and so a loop integral with $\ell$ loop momenta is UV-sensitive, going as $\left( \int \frac{d^dk}{(2\pi)^d}\right)^{\!\ell}$.

This sensitivity has a simple and physical origin: the Hamiltonian of these theories factorizes in space. In more detail, these scalar theories are the continuum version of lattice models where one has a spatial lattice $\Gamma$ with sites $i$ and a lattice spacing $a$, and the $i$th site is equipped with an independent quantum mechanics\footnote{In fact, the lattice-regularized on-site quantum mechanics model is the nearly 120-years-old Einstein model, originally used to explain the specific heat of solids at high temperatures.} with a scalar degree of freedom $\phi_i$.  Such a lattice regulator manifestly preserves the supertranslation symmetry. In an equation,
\beq
\label{E:latticeS}
	S = \sum_{i\in \Gamma} S_L[\phi_i]\,, \qquad S_L[\phi] = a^d \int du \left( \frac{1}{2}(\partial_u\phi)^2 - V(\phi)\right)\,,
\eeq
where $S_L$ is the action associated with a single lattice site. The UV-sensitivity of the free energy is now immediate, since
\beq
	F = N_{\rm sites} F_{L}\,,
\eeq
with $N_{\rm sites} = |\Gamma|$ and $F_{L}$ the free energy of the quantum mechanics living on a single site. Note that $F$ diverges in the continuum limit, since $N_{\rm sites} = \frac{V}{a^d}$ diverges as $a\to 0$ with $V$ fixed. This UV sensitivity is ultimately due to the fact that these models have a lattice-sensitive density of states. 

The ultralocality of the electric Hamiltonians is a consequence of symmetry. Continuum models with no spatial derivatives are invariant under volume preserving diffeomorphisms (VPDs). From the point of view of flat space holography this symmetry is undesirable, and it can be broken by Carroll-invariant operators with spatial derivatives as we discuss in Subsection~\ref{S:quartic}, or by coupling to a ``magnetic sector.''

However this is not the end of the story even in these ultralocal theories. The electric models in fact have far more UV sensitivity. To get a handle of the UV sensitivity we canonically normalize the kinetic term of the quantum mechanics at each lattice site in~\eqref{E:latticeS} by setting $\tilde{\phi} = a^{d/2} \phi$.  Then the action of a site is
\beq
	S_L[\tilde{\phi}] = \int du \left( \frac{1}{2}(\partial_u\tilde{\phi})^2 - V_L(\tilde{\phi})\right)\,, \qquad V_L(\tilde{\phi}) = a^d V\!\left( \frac{\tilde{\phi}}{a^{d/2}}\right)\,.
\eeq
In order for this quantum mechanics to have a finite potential $V_L$ in the continuum limit it must be the case that monomial interactions in $V$ are suppressed by powers of the lattice scale. For example, a finite quartic potential $V_L = \frac{m^2}{2}\tilde{\phi}^2 + \frac{\lambda_3}{3!}\tilde{\phi}^3 + \frac{\lambda_4}{4!}\tilde{\phi}^4$ corresponds to
\beq
\label{E:nonG1}
	V = \frac{m^2}{2}\phi^2 + \frac{\lambda_3 a^{d/2}}{3!}\phi^3 + \frac{\lambda_4 a^d}{4!}\phi^4\,.
\eeq
We remark in passing that to obtain non-suppressed non-Gaussianities in the continuum potential $V$, we need to scale couplings with inverse powers of the lattice scale, corresponding to the strong coupling limit of the on-site quantum mechanics.

The non-Gaussianities in~\eqref{E:nonG1}, suppressed by powers of the lattice scale and thereby seemingly absent in the continuum theory, comprise UV data that must be specified to obtain the density of states of a single site, an IR quantity. Equivalently the low energy and $a\to 0$ limits do not commute, which manifests UV/IR mixing.  Thus to define the continuum theory we must supply an infinite amount of UV data through the non-Gaussian part of the potential $V_L$. One way to fix that data is to impose the Carrollian analogue of conformal invariance. This, together with requiring a finite continuum limit and a polynomial potential, simply fixes the model to be free, $V_L = 0$. 

Another way to think about the UV sensitivity, present even in the free theory, is that electric operators must be normalized with suitable powers of $a$ so as to admit a smooth $a\to 0$ limit. Consider the Gaussian theory with $V_L = \frac{m^2}{2}\tilde{\phi}^2$. The mixed representation propagator of $\phi$ in the lattice-regulated theory is
\beq
\label{E:ultra1}
	\langle \phi_i(\omega)\phi_j(-\omega)\rangle = \frac{i}{\omega^2-m^2} \frac{\delta_{ij}}{a^d}\,,
\eeq
where $\delta_{ij}$ is the lattice Kronecker delta function and $\frac{\delta_{ij}}{a^d}$ becomes $\delta(x)$ in the $a\to 0$ limit. The ultralocality of~\eqref{E:ultra1} implies that a composite operator $\phi^n$ has a two-point function $\sim( \delta(x))^n$, which is badly divergent for $n>1$. These divergences can be cured by rescaling $\phi^n$ by appropriate powers of the lattice scale, i.e.~by an appropriate normalization of the operator.  There are two natural such normalizations, which we detail presently.

The first natural normalization is the scaling appropriate to study the response of a single lattice site, namely $\widetilde{O}^{(m)} = \tilde{\phi}^m$ with $\tilde{\phi} = a^{d/2} \phi$ as before. For finite $V_L$, connected correlations of these operators are local and finite on the lattice with
\beq
	\langle \widetilde{O}^{(m_1)}_{i_1} \hdots \widetilde{O}^{(m_n)}_{i_p}\rangle_{\rm conn} \sim \delta_{i_1i_2} \cdots \delta_{i_{n-1}i_n}\,.
\eeq
The second natural normalization is to take the scaling $\mathcal{O}^{(m)} = a^{\frac{d(m-1)}{2}}\phi^m= a^{-d/2} \widetilde{O}^{(m)}$ so that the two-point function has a finite, ultralocal continuum limit
\beq
\label{E:finiteOO}
	\langle \mathcal{O}^{(m)}(\omega,x) \mathcal{O}^{(m)}(-\omega,y)\rangle_{\rm conn} \sim \delta(x-y)\,.
\eeq
However higher connected moments of these operators vanish as $a\to 0$ (again when $V_L$ is finite), e.g.\ 
\begin{equation*}
	\langle \mathcal{O}^{(m_1)}(\omega,x)\mathcal{O}^{(m_2)}(\omega',y) \mathcal{O}^{(m_3)}(-\omega-\omega',z)\rangle_{\rm conn} \sim a^{d/2} \delta(x-y)\delta(y-z) \to 0\,.
\end{equation*}
Relatedly, for finite $V_L$ there is no normalization of composite operators that is both finite in the continuum limit and furnishes non-Gaussianities correlations. In summary, having a finite on-site potential implies that the continuum theory has only Gaussian correlations.

These above features are reminiscent of the behavior of correlation functions for large $N$ gauge theories in the 't Hooft limit. There, $n$-point connected correlation functions of normalized single trace operators are of order $N^{2-n}$ and so correlations are Gaussian to $O(N^0)$. Indeed, in the lattice models above, the connected $n$-point function of normalized operators is of order $\left( a^{-d/2}\right)^{2-n}$ so that $a^{-d/2}$ is playing the role of a large $N$ parameter. Equivalently, we can think of the continuum limit as a classical one with $a^{d/2}$ corresponding to an effective Planck's constant $\hbar$. For this reason we can think of our models as being theories of generalized free fields, one for each normalized operator.

While connected $n$-point functions of normalized operators $\mathcal{O}$ vanish in the continuum limit for $n>2$, one-point functions generically diverge as $a^{-d/2}$. This immediately follows from $O(a^0)$ expectation values for on-site operators $\langle \widetilde{O}\rangle$ and can be seen concretely for a model with an on-site quartic potential. With~\eqref{E:nonG1} the saddle-point value of $\phi$ vanishes. However had we included a finite linear term in the on-site potential $V_L$, namely $V\supset \lambda_1 \tilde{\phi}$, this would have generated a large term in the continuum potential $V\supset \lambda_1 a^{-d/2} \phi$ along with a large saddle-point value for $\phi$ of $ O(a^{-d/2})$. Indeed, even in the absence of such a linear term, there is a large $O(a^{-d/2})$ tadpole generated by the cubic interaction. 

We are now in a position to make sense of the UV-sensitivities of loop integrals. With a hypercubic lattice regulator we have $\int \frac{d^dk}{(2\pi)^d} = \frac{1}{a^d}$ since the momenta live on a torus with $k \sim k + \frac{2\pi}{a} e$ for $e$ a unit lattice vector. In the Gaussian theory the two-point function of $\mathcal{O}^{(n)}$ is an $(n-1)$-loop diagram with $n$ tree-level propagators connecting the two insertions. The loop integrand is independent of spatial momenta, and so the momentum dependence of two-point function is
\beq
	\langle \mathcal{O}^{(n)}(\omega,k)\mathcal{O}^{(n)}(-\omega,-k)\rangle_{\rm conn} \sim \left( a^{\frac{d(n-1)}{2}}\right)^2 \left( \int\frac{d^dq}{(2\pi)^d}\right)^{n-1} = O(a^0)\,, 
\eeq
where the prefactor comes from the normalization of $\mathcal{O}^{(n)}$. The result is finite and independent of the external momentum in the $a\to 0$ limit. Upon Fourier transform back to configuration space one recovers $\langle \mathcal{O}^{(n)}(\omega,x)\mathcal{O}^{(n)}(-\omega,y)\rangle \sim \delta(x-y)$.

So far we have seen that simple electric models defined through a certain continuum limit of the theory on a lattice are consistent. But can we reliably compute properties of the continuum theory without introducing a lattice regulator? Thankfully the answer is ``yes'' when the on-site quantum mechanics admits a diagrammatic expansion. However, as we will see in the next Section, the answer can be ``no'' when there are non-perturbative effects that are sensitive to the details of the lattice, like winding modes for a compact scalar or for gauge theory.

Here we develop this continuum prescription. The basic idea is to back slightly off of the continuum limit and work on a hypercubic lattice of infinitesimal but nonzero lattice spacing $a$. Consider an on-site model which admits a diagrammatic expansion, like~\eqref{E:nonG1}. Then, as mentioned above, the tree-level propagator for $\phi$ is
\beq
	\langle \phi_i(\omega)\phi_j(-\omega)\rangle_0 = \frac{i}{\omega^2-m^2} \frac{\delta_{ij}}{a^d}\,,
\eeq
and there is a cubic vertex $i\lambda_3 a^{d/2}$ as well as a quartic vertex $i\lambda_4 a^d$. A similar analysis of this theory was performed in~\cite{Banerjee:2023jpi}.  While we could work in momentum space, which is rendered periodic on a lattice, it is more convenient to work in a mixed frequency-position representation. Consider the $n$-point function of normalized composite operators $\mathcal{O}^{(m_{\alpha})} = a^{\frac{d(m_{\alpha}-1)}{2}}\phi^{m_\alpha}$ with $a=1,2,...,n$. In perturbation theory one simply writes down all diagrams that contribute to a process with $n$ insertions of those operators, i.e.\ with $m_\alpha$ $\phi$-lines emanating from $n$ sources, each of which is weighted by a suitable power of $a$. Loop integrals include an integral over loop frequencies (which becomes a sum over Matsubara modes at finite temperature in the imaginary time formalism), and a sum over the position of internal vertices. For an $n$-point vertex that sum takes the form $a^d\sum_{z \in \Gamma}\delta_{i_1,z} \cdots \delta_{i_n, z}$ where the $i_k$ each refer to a site index of a propagator which is to be attached to the vertex.   The Kronecker deltas sets those sites $i_k$ to be coincident at $z$, and $z$ is summed over the lattice.

So far we have not done anything non-trivial. This is merely lattice perturbation theory. However, a bit of experimentation and thought reveals that the diagrams for connected $n$-point functions scale as $a^{\frac{d(n-2)}{2}}$. This is precisely the same scaling we arrived at above, and it implies that tadpoles generically diverge in the continuum limit while connected two-point functions remain finite. For example, consider the one-loop tadpole for $\phi$, as well as the one-loop corrections to the propagator. The tadpole is 
\begin{center}
\vspace{-.05in}
\includegraphics[width=0.7in]{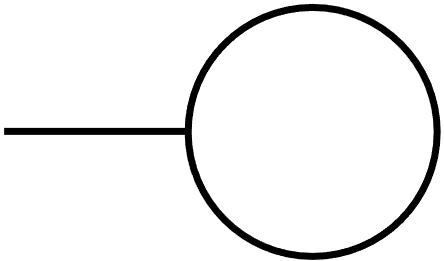}
\vspace{-.05in}
\end{center}
which evaluates to
\beq
 	\langle \phi_i(\omega)\rangle =  \left( \frac{i}{\omega^2-m^2} \frac{\delta_{ij}}{a^d} \right)\left( a^d\sum_{j,k,l\in\Gamma}  (i \lambda_3 a^{d/2})\delta_{jk}\delta_{kl}\right) \frac{1}{2} \int \frac{d\omega'}{2\pi}\frac{i}{\omega'^2-m^2}\frac{\delta_{kl}}{a^d} (2\pi\delta(\omega))  + O(\lambda_3\lambda_4,\lambda_3^3)\,.
\eeq
The first term corresponds to the propagator for the dangling $\phi$ propagator; the second to the interaction vertex including the sum over locations where the interaction may take place; the $\frac{1}{2}$ is a symmetry factor; and the rest comes from the loop. The sum over lattice sites collapses thanks to ultralocality. Fourier transforming to real time the result is
\beq
\label{E:electricTadpole}
	\langle \phi\rangle = - \frac{\lambda_3}{4|m|^3} \,a^{-d/2} + O(\lambda_3\lambda_4,\lambda_3^3)\,.
\eeq
The loop-corrected propagator is
\begin{center}
\vspace{-.05in}
\includegraphics[width=3.5in]{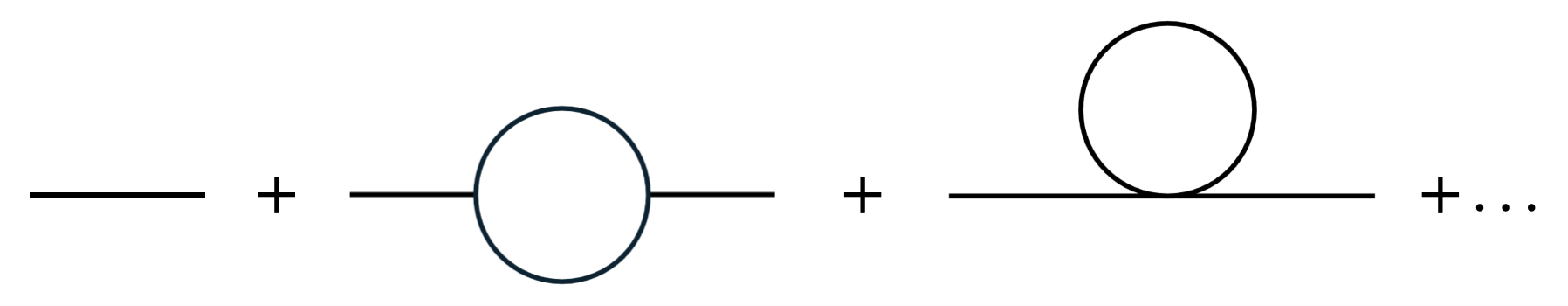}
\vspace{-.05in}
\end{center}
which after a similar collapse of sums over lattice sites reads
\begin{align}
	\langle \phi_i(\omega)\phi_j(-\omega)\rangle_{\rm conn} & = \langle \phi_i(\omega)\phi_j(-\omega)\rangle_0 
	\\
	\nonumber
	&\hspace{-.4in} + \left( \frac{i}{\omega^2-m^2}\right)^2 \frac{\delta_{ij}}{a^d} \frac{1}{2}\int \frac{d\omega'}{2\pi}\left( (i\lambda_3)^2 \frac{i}{(\omega-\omega')^2-m^2}\frac{i}{\omega'^2-m^2} + i \lambda_4 \frac{i}{\omega'^2-m^2}\right)+ \cdots\,.
\end{align}
Here the dots indicate two- and higher-loop corrections. Evaluating the integral we have
\begin{align}
\begin{split}
	\langle \phi_i(\omega)\phi_j(-\omega)\rangle_{\rm conn} = \langle \phi_i(\omega)\phi_j(-\omega)\rangle_0- \frac{i}{(\omega^2-m^2)^2}\frac{\delta_{ij}}{a^d} \left(  \frac{\lambda_3^2}{4\omega(|m|\omega+2\omega^2)}+ \frac{\lambda_4}{4|m|^3}\right) + \cdots\,.
\end{split}
\end{align}
The tadpole diverges in the continuum limit as expected while the two-point function is smooth, with $\frac{\delta_{ij}}{a^d}\to \delta(x-y)$. 

The results above, and more general diagrams of the lattice theory, can be obtained by a simple mnemonic where we work in terms of continuum positions $x$ rather than lattice sites, which was also discussed in~\cite{Banerjee:2023jpi}.  In fact this mnemonic nearly coincides with the na\"{i}ve path integral treatment. In this prescription we take the mixed frequency-position propagators to be
\beq
	\langle \phi(\omega,x)\phi(-\omega,y)\rangle_0 = \frac{i}{\omega^2-m^2} \delta(x-y)\,,
\eeq
we integrate over the continuous position of interaction vertices, and we retain the lattice spacing in the cubic and quartic vertices which we take to be $i\lambda_3a^{d/2}$ and $i\lambda_4a^d$ respectively. Where necessary we also take $\delta(0) = a^{-d}$. For the tadpole $\langle \phi \rangle$ at one-loop, this gives (the P is for ``prescription'')
\begin{align}
\begin{split}
	\langle \phi \rangle_{\rm P}& = \left( i \lambda_3a^{d/2} \int d^dy d^dz d^dw \delta(y-z)\delta(z-w) \right) \left( \frac{i}{\omega^2-m^2}\delta(x-y)\right) 
	\\
	 & \qquad \qquad \times \frac{1}{2} \int \frac{d\omega'}{2\pi} \frac{i}{\omega'^2-m^2} \delta(z-w) (2\pi\delta(\omega))
	 \\
	  &=  - \frac{\lambda_3}{4|m|^3} (2\pi\delta(\omega)) (a^{d/2}\delta(0)) 
	  \\
	   & \to  - \frac{\lambda_3}{4|m|^3}(2\pi\delta(\omega)) a^{-d/2}\,,
\end{split}
\end{align}
which coincides with~\eqref{E:electricTadpole}. Similarly, if we compute correlations of composite operators, then we include explicit powers of the lattice scale in the definition of those operators.

This prescription will be especially useful for us in Section~\ref{S:electromagnetic} when we study interactions between electric and ``magnetic'' degrees of freedom and a Carrollian version of scalar QED.

\subsection{Higher derivatives}
\label{S:quartic}

All the electric theories considered above are trivial in the sense that the Hamiltonian spatially factorizes. As we mentioned, the triviality is ultimately a consequence of symmetry, namely invariance under volume preserving diffeomorphisms in the continuum limit. At the two derivative level there are no Carroll-invariant deformations that break VPD invariance. What about higher derivative deformations?

Classically there are deformations with four or more derivatives which are invariant under supertranslations but not VPDs~\cite{Baig:2023yaz} (see also~\cite{Ecker:2024lur}). These deformations include operators with spatial derivatives and at least two real scalar fields. To see the invariance, we note that $\partial_i \rightarrow \partial_i + (\partial_if) \partial_u$ under supertranslations $u \rightarrow u + f(x)$, and so a term like $\partial_u\phi_1 \partial_i \phi_2 - \partial_i \phi_1 \partial_u\phi_2$ is in fact invariant. Thus the simplest example of a supertranslation-but-not-VPD invariant field theory involves a single complex scalar field $\phi$, and has the action\footnote{We could also include a single time derivative term $i \int du \,d^dx \,\bar{\phi}\partial_u \phi$.}
\beq
 \label{eq:interacting_scalar}
    S = \int dud^dx \left(\frac{1}{2}|\partial_u\phi|^2 - V(|\phi|^2)+g \left|\partial_u\bar{\phi} \partial_i \phi - \partial_i \bar{\phi}\partial_u\phi  \right|^2 \right)\,.
\eeq
The quartic interaction preserves a $U(1)$ symmetry that rotates $\phi$, so we write a potential that does the same. Due to the spatial derivatives the Hamiltonian no longer spatially factorizes, even though it is supertranslation invariant and so has a locally conserved energy. The above theory can be thought of as the Carrollian limit of a relativistic scalar field theory with certain higher-derivative interactions~\cite{Kasikci:2023tvs}. We also note that the interaction term explicitly breaks the VPD symmetry of the action to just supertranslations.

There are related deformations of electromagnetic theories. Consider first taking the $c\to 0$ limit for pure electromagnetism,
\begin{equation}
    S = \frac{c}{4e^2} \int du d^dx F^2 = \frac{1}{2e^2 c} \int du dx \left(E^2 -c^2 B^2 \right)  \to  \frac{1}{2e^2} \int du d^dx \,E^2 \,,
\end{equation}
where we have taken the $c\to 0$ limit after either rescaling $A_{\mu} \to \sqrt{c}A_{\mu}$ or the gauge coupling $e^2 \to e^2/c$, one can include higher-order terms hoping that they have non-trivial Carrollian limits. In $2+1$ there are no other scalar invariants, and so the Carrollian limit is trivial. However, in $3+1$ there is an additional scalar invariant $\det(F)  = \left(2F^4 - (F^2)^2\right)/8  =  (E \cdot B)^2/c^2$, and so there is a non-trivial Carrollian limit,
\begin{equation}
\label{E:quarticEM}
    S =  \int du\,d^3x \left( \frac{1}{2e^2}E^2 + g \left(E \cdot B \right)^2 \right).
\end{equation}
Let us verify that this theory is indeed supertranslation invariant.  The generalization of the field transformations under the global Carrollian symmetries from \cite{Duval:2014uoa,Henneaux:2021yzg} to supertranslations are rather straightforward, with $\vec{E}$ invariant while $\vec{B}$ transforms as $\vec{B} \rightarrow \vec{B} + (\nabla f) \times \vec{E}$. We can also work with the gauge potential directly, which transforms like the derivatives under supertranslations:
\begin{equation}
    A_u \rightarrow A_u\,, \qquad  A_i \rightarrow A_i + (\partial_i f)A_u\,.
\end{equation}

In higher dimensions the analogous leading deformation is
\begin{equation}
    \delta S = g \int du \int_{M^d} (E \wedge B)^2,
\end{equation}
with the transformation law for the magnetic two-form being $B \rightarrow B + df \wedge E$. 

The quantization of \eqref{eq:interacting_scalar} is difficult. One strategy is to treat the interaction as if it is perturbatively small. This is subtle, as the interaction vertex is quadratic in spatial momenta but the tree-level propagator is momentum-independent, leading to divergent loop integrals where the integrand is polynomial in momenta. Given the discussion in the last Subsection, a natural regularization would be to put the theory on a lattice, but at the present time it is not clear how to do so in such a way as to preserve the supertranslation symmetry.

One way to proceed is to consider a generalization of~\eqref{eq:interacting_scalar} with $N \gg 1$ complex fields $\phi^a$, where we impose a $U(N)$ global symmetry.  Such a theory can be solved in the leading large $N$ limit by solving the Dyson equations for the scalar propagator and its self-energy; this approach has been used to great effect in Chern-Simons-matter theories (see e.g.~\cite{Giombi:2011kc}), the SYK model (e.g.~\cite{Rosenhaus:2018dtp}), and models with dipole symmetry~\cite{Jensen:2022ibn}. For concreteness, suppose our theory has a quartic potential $V(|\phi|^2) =m^2|\phi|^2 + \lambda_4 |\phi|^4$.  Using the results of e.g.~\cite{Jensen:2022ibn} the Dyson equations for the Euclidean propagator $G(\omega,k) =\frac{1}{N} \langle\bar{\phi}^a(\omega,k) \phi^a(-\omega,-k) \rangle$, the self energy $\Sigma(\omega,k)$, and the $U(1)$ condensate $\sigma$, are
\begin{equation}
    \begin{aligned}
        G(\omega ,k) &= \frac{1}{\omega^2 + \Sigma(\omega,k)} + \frac{|\sigma|^2}{N} (2\pi)^{d+1} \delta(\omega) \delta^{(d)}(\vec{k})\,,
        \\
        \Sigma(\omega,k)&= m^2 + 2\int \frac{d\omega' d^dk'}{(2\pi)^{d+1}}\left( \lambda_4+ g\left(\omega k' - k \omega' \right)^2\right)G(\omega',k')\,, 
    \end{aligned}
\end{equation}
along with the condition $\sigma\,\Sigma(\omega =0, k=0) = 0$. One can make a self-consistent rotationally invariant anstatz for the self energy
\begin{equation}
    \Sigma(\omega,k) = a_0 + a_1 k^2 + a_2 \omega^2\,,
\end{equation}
which results in a finite set of coupled equations for the undetermined parameters $\{a_0,a_1,a_2,\sigma\}$,
\begin{equation} \label{eq:Dyson}
    \begin{aligned}
        a_0 &= m^2 +2\lambda_4  \int \frac{d\omega d^dk}{(2\pi)^{d+1}}\frac{1}{\omega^2 + \Sigma(\omega,k)} + 2\lambda_4 \frac{|\sigma|^2}{N}\,,
        \\
        a_1 &= 2g \int \frac{d\omega d^dk}{(2\pi)^{d+1}}\frac{\omega^2}{\omega^2 + \Sigma(\omega,k)}\,,
        \\
        a_2 &= 2g \int \frac{d\omega d^dk}{(2\pi)^{d+1}}\frac{k^2}{\omega^2 + \Sigma(\omega,k)} \,,
    \end{aligned}
\end{equation}
along with the condition $a_0 \sigma = 0$.

The integrals in the above Dyson equations~\eqref{eq:Dyson} diverge and so need to be regulated, ideally in such a way as to preserve the supertranslation symmetry. On general grounds one expects the dominant solution to have $a_i\neq 0$, in which case resumming the quartic higher derivative interaction would lead to a two-point function that has a non-trivial dependence on both $\omega$ and $k$. This would in turn imply the spontaneous breaking of not only supertranslations, but even the Carroll boost symmetry $u \to u + \vec{a}\cdot \vec{x}$.

\section{Carrollian electromagnetism}
\label{S:EM}

In this Section we take a brief foray into the study of seemingly gapless electric theories. Our primary goal is to understand the Carrollian version of electromagnetism that we mentioned briefly in the last Subsection, described by the action
\beq
\label{E:electro1}
	S = \frac{1}{2g^2}\int du d^dx \,E^2\,, \qquad E_i = \partial_u A_i - \partial_i A_u\,,
\eeq
that follows from the na\"{i}ve $c\to 0$ limit of ordinary electromagnetism. Crucially we will take $A_{\mu}$ to be a $U(1)$ gauge field rather than a $\mathbb{R}$ gauge field. In $A_u = 0$ gauge we have $d$ decoupled electric degrees of freedom $A_i$ subject to the Gauss' Law constraint. As we will see, this seemingly simple Gaussian theory is rather subtle. To proceed we will replace continuous space by a lattice as in the previous Section. In the regulated theory the vector potential $A_i$ becomes a set of phase variables $e^{i\varphi}$ living on the links of the lattice. We then have $d$ copies of the lattice-regulated theory of a compact scalar $\varphi$. To set the stage we first study the lattice-regulated compact scalar theory in detail, and then go on to solve Carrollian electromagnetism.

\subsection{Warmup: compact scalar}

Let us consider the electric theory of a compact, non-interacting scalar $\varphi$ described by the action
\beq
	S = \frac{R^2}{2} \int du\, d^dx \,(\partial_u \varphi)^2\,, \qquad \varphi \sim \varphi + 2\pi\,.
\eeq
This is perhaps the simplest Carrollian theory we can study.  One way of arriving at it is to take the $c\to 0$ limit of the relativistic action for a compact scalar $\varphi$, namely $-\frac{\tilde{R}^2}{2} \int d^{d+1}x \,(\partial\varphi)^2$, together with an identification $\tilde{R}^2 = R^2/c$ so that the theory admits a smooth $c\to 0$ limit. Na\"{i}vely one might think that the Carrollian correlators are simple $c \to 0$ limits of those of the relativistic theory, but we will see that this is not strictly true. The periodicity of $\phi$ generates a scale in $d\neq 1$ dimensions, so despite the absence of a mass-like term this is a scale-non-invariant theory in general dimension, unless we take $R\to \infty$ in which case we have a Carrollian CFT. 

By now we know the right way to analyze the Carrollian theory, namely to regulate its UV sensitivity by placing it on a spatial lattice. However to illustrate the importance of such a careful treatment let us first perform a na\"{i}ve continuum path integral quantization. 

\subsubsection{Na\"{i}ve quantization}

Here we consider the na\"{i}ve continuum path integral quantization of the Carrollian compact scalar, purely for pedagogical purposes.  This treatment gives rise to incorrect answers, for subtle reasons we will explain shortly.

Suppose we want to compute the thermal partition function using the imaginary time formalism at inverse temperature $\beta$ and where we compactify space to be a rectangular torus whose sides have lengths $L_i$ and volume $V = L_1\cdots L_d$. Because $\varphi$ is compact and we are working on a Euclidean torus we have field configurations where $\varphi$ winds around the cycles of the torus, and we must sum over all possible windings. Allowing for winding number $m$ around the thermal circle $\tau = i u$, spatial winding numbers $n_i$, and decomposing the fluctuations around the winding configuration into Fourier modes, we parameterize $\varphi$ as
\beq
	\varphi = \frac{2\pi m}{\beta} \tau + \sum_{i=1}^d\frac{2\pi n_i x^i}{L_i} + \frac{1}{\sqrt{\beta V}} \sum_{\omega,k} e^{i \omega \tau + i k\cdot x} \delta\varphi(\omega,k)\,, \qquad \delta \varphi(\omega,k)^* = \delta\varphi(-\omega,-k)\,,
\eeq
and $ \omega = \frac{2\pi \mathbb{Z}}{\beta}$, $k_i = \frac{2\pi \mathbb{Z}}{L_i}$. The Euclidean action of this configuration is
\beq
	S = \frac{R^2V}{2\beta} \left(2\pi m \right)^2 +\frac{R^2}{2} \sum_{\omega,k}\omega^2|\delta\varphi(\omega,k)|^2\,.
\eeq
The frequency zero modes drop out and the integration over them produces an IR divergent prefactor to the partition function that can be interpreted as a field space delta function $\delta[0]$. Dualizing the sum over windings $m$ into one over ``dual'' windings $\tilde{m}$ using the Poisson summation formula and using\footnote{To derive this expression we start with the one-loop determinant for an oscillator, $\prod_{\omega} \frac{1}{\sqrt{\omega^2+m^2}} = \frac{1}{2\sinh\left( \frac{\beta m}{2}\right)}$. We then delete the Fourier zero mode by multiplying by $m$, and then take the $m\to 0$ limit.} 
\beq
	\prod_{\omega\neq 0} \frac{1}{|\omega|} = \frac{1}{\beta}\,,
\eeq
we have
\beq
\label{E:result0}
	Z_{\text{na\"{i}ve}} =\delta[0] \sum_{\tilde{m},n_i\in \mathbb{Z}} e^{-\frac{\beta \tilde{m}^2}{2R^2V}} \prod_{k\neq 0} \frac{1}{\beta}\,.
\eeq
This expression has three sources of divergence. One is the aforementioned infrared divergent prefactor $\delta[0]$; another is the divergence from the sum over spatial windings, which should be interpreted as UV sensitivity since the spatial windings can have arbitrarily large gradients and yet cost zero energy; the third is the one-loop determinant $\prod_{k\neq 0} \frac{1}{\beta}$, whose logarithm goes like $\int \frac{d^dk}{(2\pi)^d}$ and thus is UV-sensitive, although in a manner we have already encountered and which is easily regulated with a spatial lattice.

It is instructive to compare the result~\eqref{E:result0} with that of a relativistic compact scalar as $c\to 0$. The partition sum for that theory at finite $c$ is
\beq
	Z_{\rm rel} = \sum_{\tilde{m},n_i\in\mathbb{Z}} e^{-\frac{\beta \tilde{m}^2}{2R^2V} - \sum_{i=1}^d \frac{\beta (2\pi)^2 VR^2 c^2 }{2 L_i^2}} \prod_{k\neq 0} \frac{1}{2\sinh\left( \frac{\beta c|k|}{2}\right)}\,.
\eeq
The first term in the exponential refers to the dual of the sum over windings around the thermal circle, and the second term to windings around the spatial circles. It is straightforward to map $Z_{\rm rel}$ to $Z_{\text{na\"{i}ve}}$. The winding modes of one become those of the other, with the spatial windings becoming gapless in the $c\to 0$ limit. Next, it is useful to write the oscillator partition function $\frac{1}{2\sinh\left( \frac{\beta c |k|}{2}\right)}$ as $(c|k|)^{-1}\left(  \frac{c |k|}{2\sinh\left( \frac{\beta c|k|}{2}\right)}\right)$. The first part comes from the Fourier frequency zero mode and diverges in the $c\to 0$ limit, while the second becomes $\frac{1}{\beta}$ as $c\to 0$. Now including the infinite product over $k\neq 0$, the frequency zero mode contributions become $\delta[0]$ while the rest becomes the one-loop prefactor $\prod_{k\neq 0}\frac{1}{\beta}$. 

In sum, the $c\to 0$ limit of the relativistic compact scalar becomes the na\"{i}ve path integral treatment of the Carrollian compact scalar. However, this agreement is misleading, and the na\"{i}ve answer $Z_{\text{na\"{i}ve}}$ for the partition sum is in fact hopelessly incorrect. 

\subsubsection{Correct quantization}

What is the correct answer, and what went wrong above?  To answer the first question let us regulate the Carrollian theory by placing it on a hypercubic spatial lattice $\Gamma$ of lattice spacing $a\ll L_i$. The real-time action is
\beq
	S = \frac{a^dR^2}{2} \sum_{i\in \Gamma} \int du\,(\partial_u \varphi_i)^2\,,
\eeq
where $i$ labels a lattice site. This is the theory of independent rotor degrees of freedom at each lattice site. The correct partition sum is then
\beq
	Z_{\rm correct} = Z_L^{N_{\rm sites}}\,, \qquad Z_L = \sum_{\tilde{m}\in\mathbb{Z}} e^{-\frac{\beta \tilde{m}^2}{2a^dR^2}}\,.
\eeq
where $Z_L$ is the partition function for a single site.  The physics of this answer is completely different from that of the na\"{i}ve path integral. The correct result describes independent rotor degrees of freedom at each site with a spectrum of energies $E_{\tilde{m} }= \frac{\tilde{m}^2}{2a^dR^2}$. In the limit of $R$ finite and $a\to 0$ the low-energy spectrum contains a single state, where each rotor is in its ground state. As such the theory is far from being gapless, which one may have expected from the absence of a mass-like term in the action. 

We can also study simple correlation functions. The rotor has an independently conserved integer-valued charge at each site, the momentum conjugate to $\phi_i$, namely $\widetilde{Q}_i = a^d \partial_u\phi_i$. Due to the conservation of charge and the gap its zero-temperature two-point function vanishes,
\beq
	\langle \widetilde{Q}_i(u) \widetilde{Q}_j(0)\rangle = 0\,.
\eeq
(This result actually agrees with the prediction of the na\"{i}ve continuum path integral.) We can also define a normalized version of the operator $e^{i n\phi_i}$ with $n\in\mathbb{Z}$, namely $\mathcal{O}^{(n)}_i = a^{-d/2} e^{in \phi_i}$, whose two-point function in the zero-temperature limit reads
\beq
\label{E:compactVertex}
	\langle \mathcal{O}^{(n)}_i(u)\mathcal{O}^{(m)}_j(0)\rangle = e^{-\frac{i n^2u}{2a^dR^2}}\delta_{nm}\frac{\delta_{ij}}{a^d}\,.
\eeq
While the position-dependence admits a continuum limit with $\frac{\delta_{ij}}{a^d}\to \delta(x-y)$, the oscillatory prefactor does not unless we scale $R\sim a^{-d/2}$ as $a\to 0$.  Let us study that scaling limit, i.e.\ we take $R^2 \to a^{-d} R^2$. Now the on-site theory is that of a rotor with a finite moment of inertia $R^2$ and so also a finite on-site spectrum $E_{\tilde{m}} = \frac{\tilde{m}^2}{2R^2}$. The two-point function is also smooth with the result
\beq
	\langle \mathcal{O}^{(n)}(u,x)\mathcal{O}^{(m)}(0,y)\rangle = e^{-\frac{in^2u}{2R^2}}\delta_{n+m,0}\,\delta(x-y)\,.
\eeq
As for the simple electric theories we studied in Section~\ref{S:electric2}, connected $n$-point functions of the $\mathcal{O}^{(m)}$'s scale as $a^{\frac{d(n-2)}{2}}$ and so have only Gaussian correlations as $a\to 0$.

We are now in a position to understand where our na\"{i}ve path integral (and $c\to 0$) analysis went awry. Recall our result at the end of Subsection~\ref{S:electric2}: for simple electric theories with a single saddle point, lattice perturbation theory around the saddle can be described as a prescription directly in the continuum. However, the lattice version of the compact scalar has saddles that have no continuum analogue, namely configurations where the scalar winds around the thermal circle with different winding numbers depending on the lattice site. In an ordinary lattice theory with spatial gradient terms these configurations carry infinite action in the continuum limit and so their contribution is suppressed. In the Carrollian theory those gradients are absent, these configurations are unsuppressed, and their effect is completely missed in the na\"{i}ve path integral treatment.

Another way of stating the problem is that our na\"{i}ve path integral did not sum over field configurations which are smooth in time but discontinuous in space, in particular configurations where the winding number is a discontinuous function of spatial position. Discontinuous field configurations are also known to play an important role in exotic field theories with subsystem symmetry~\cite{Seiberg:2020bhn,Seiberg:2020wsg}. 

The above lessons immediately carry over for the Carrollian version of nonlinear sigma models, $\int du \,d^dx \frac{g_{ab}(X)}{2}\partial_u X^a \partial_u X^b$. It is natural to define these models so that the on-site quantum mechanics is finite in the continuum limit, which for a compact sigma model will lead to a discrete spectrum which, from a path integral point of view, arises from configurations winding around the thermal circle discontinuously as a function of position.

While the compact scalar differs significantly from its na\"{i}ve path integral treatment, the noncompact massless scalar $\varphi$ described by the action $S = \int du \,d^dx \frac{1}{2}(\partial_u \varphi)^2$ can be treated in a more or less conventional way. This model has a family of saddles labeled by the zero mode of the scalar, and fluctuations around each can be described by a continuum diagrammatic prescription. This model is gapless with a continuous spectrum of on-site energy eigenstates with energies $E(p) = \frac{p^2}{2a^d}$. This noncompact scalar theory, while quite simple, has the virtue of being the only conformally-invariant version of a Carrollian theory to appear in this Section or the last one.

\subsection{Electromagnetism}
\label{S:pureEM}

Now we turn our attention to Carrollian electromagnetism. We wish to place the theory~\eqref{E:electro1} on a spatial hypercubic lattice $\Gamma$ with lattice spacing $a$. At each site $i\in \Gamma$ we have $A_{u,i}$, and on each link $\ell$ connecting the site $i$ to the site $j$ there is a compact scalar $\varphi$. Given a link $\ell$ connecting $i$ and $i+e_m$, we let $\varphi_{i\to i+e_m}$ refer to $\varphi$ on that link, while $\varphi_{i+e_m\to i}$ refers to $-\varphi$ there. The desired action is
\beq
\label{E:desiredaction1}
	S = \frac{a^d}{2g^2} \sum_{i\in \Gamma} \sum_{m=1}^d E_{i,m}^2\,, \qquad E_{i,m} = \frac{1}{a}\left(   \partial_u \varphi_{i\to i+e_m}- A_{u,i+e_m}+ A_{u,i}\right)\,.
\eeq
Equivalently the double sum is a single sum over lattice links with a preferred orientation. Gauge transformations act on the lattice sites via $U(1)$ transformation parameters $\Lambda_i \sim \Lambda_i+2\pi$ and under them the degrees of freedom $A_{u,i}$ and $\varphi_{i\to i+e_m}$ transform as
\beq
	A_{u,i}\to A_{u,i} + \partial_u \Lambda_i\,, \qquad \varphi_{i\to i+e_m} \to \varphi_{i\to i+e_m} +\Lambda_{i+e_m}- \Lambda_i\,.
\eeq
In the continuum limit $\frac{\varphi_{i\to i+e_m}}{a}$ becomes the vector potential $A_m(u,x)$ with $x$ the position corresponding to the lattice site $i$.

We can quantize~\eqref{E:desiredaction1} immediately in $A_u = 0$ gauge. In the Hamiltonian formalism we have a model of decoupled compact scalars $\varphi$ living on each link subject to the Gauss' Law constraint. In the pre-constrained system the on-link electric field is quantized and conserved with $E \in \frac{g^2}{a^{d-1}}\mathbb{Z}$; the Hilbert space is spanned by simultaneous eigenstates of the electric field, $\bigotimes_{\ell\in \Gamma_{\ell}} |E_{\ell}\rangle$ with $\ell$ a link and $\Gamma_{\ell}$ the set of links; and the total energy is $H = \frac{a^d}{2g^2}\sum_{\ell \in \Gamma_L} E_{\ell}^2$. The Gauss' Law constraint states that there is zero total electric field entering each vertex on the lattice. 

On a one-dimensional periodic lattice the only way to solve the Gauss' Law constraint is for the electric field to be constant. This results in the spectrum of energies $E_n = L\frac{g^2n^2}{2}$ with $n\in\mathbb{Z}$ and $L = N a$ the physical length of the lattice with $N$ the number of lattice sites. But this is nothing more than the standard result for the spectrum of 1d electromagnetism.

\begin{figure}[t]
\begin{center}
\includegraphics[width=1.5in]{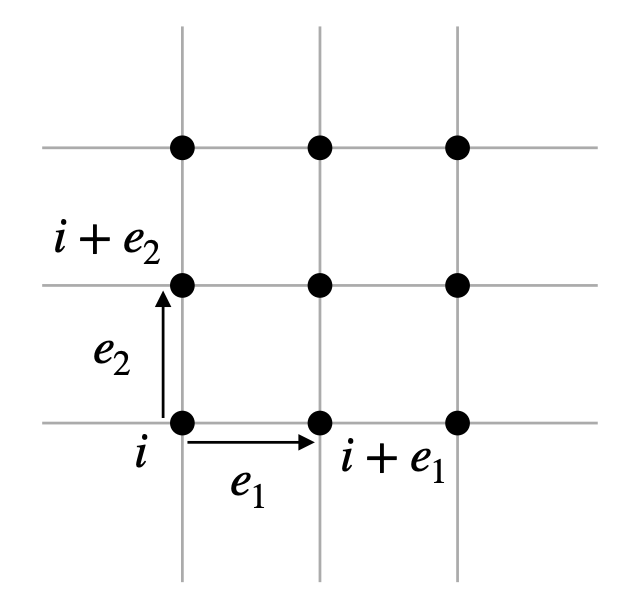} \qquad \includegraphics[width=1.5in]{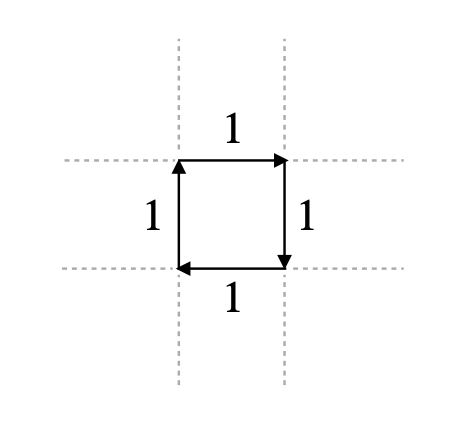}
\caption{ \label{F:lattice} (Left) A plane of a hypercubic lattice spanned by the lattice vectors $e_1$ and $e_2$. Each site is equipped with the degree of freedom $A_u$ and each link with a compact scalar $\varphi$. (Right) The elementary excitation of the electric field, with one unit of electric field running around the edges of a plaquette.}
\end{center}
\end{figure}

In more than one spatial dimension the Gauss' Law constraint allows local fluctuations of the electric field. The elementary excitation is concentrated on the edges of a spatial plaquette, with a single unit of electric field circulating around its edges. See Fig.~\ref{F:lattice}. The most general arrangement satisfying the Gauss' Law is a superposition of these states, and the energy of each state scales as $\frac{g^2}{a^{d-2}}$. 

In our treatment of scalar QED we will see that it is useful to scale the effective coupling with the lattice spacing as $g_{\rm eff} = a^{d/2} g$ with $g$ held finite as $a\to 0$. With that scaling the energy of excitations is of order $(a g)^2$, the spatial area of the plaquette, leading to a gapless theory in the continuum. 

Because the electric field is conserved, up to contact terms in time the zero-temperature two-point function of $E$ then vanishes, just like the two-point function of the charge operator of the compact scalar,
\beq
	\langle E_{i,m}(u) E_{j,n}(0)\rangle = 0\,.
\eeq
A more interesting observable is a Wilson loop, either an entirely spatial loop, i.e.\ an extended operator, or one partially extended in time, which creates a non-trivial state with a charge-anticharge pair. Spatial Wilson loops are easy to compute: given an oriented curve $\mathcal{C}$ at fixed time, the Wilson loop for charge $q\in\mathbb{Z}$ is a product over links in $\mathcal{C}$,
\beq
	\mathcal{W}_{\mathcal{C}}^{(q)} = \prod_{\ell\in\mathcal{C}} e^{ \pm i q \varphi_{\ell}}\,,
\eeq
where the $\pm$ refers to whether the orientation of $\varphi$ is aligned or anti-aligned with the orientation of the curve $\mathcal{C}$. Before imposing the Gauss' Law constraint we have seen that we can specify the Hilbert space in terms of states of the electric field on the links. Given a link $\ell$ we can parameterize those states as $|n\rangle$ with $n$ the number of units of electric field. In this basis the operator $e^{i q \varphi_{\ell}}$ is akin to the operator $e^{iq \varphi}$ of the compact scalar, with $\langle m | e^{i q \varphi_{\ell}} |n\rangle = \delta_{m,n+q}$. Then acting on the vacuum with $\mathcal{W}_{\mathcal{C}}$ produces with amplitude unity a state in which $q$ units of electric field wind around the curve $\mathcal{C}$.\footnote{The Hilbert space can be generated by acting on the ground state with products of Wilson loops.} This state has energy $E_{\mathcal{C}}^{(q)} = \frac{g^2q^2}{2a^{d-2}}N_{\mathcal{C}}$ with $N_{\mathcal{C}}$ the number of links in $\mathcal{C}$. Note that the energy of this state scales like the length of the curve. In the scaling limit with $g_{\rm eff} = a^{d/2} g$ this energy goes as $E \sim a g^2 L_{\mathcal{C}}$ with $L$ the physical length.

\begin{figure}[t]
\begin{center}
\includegraphics[width=2in]{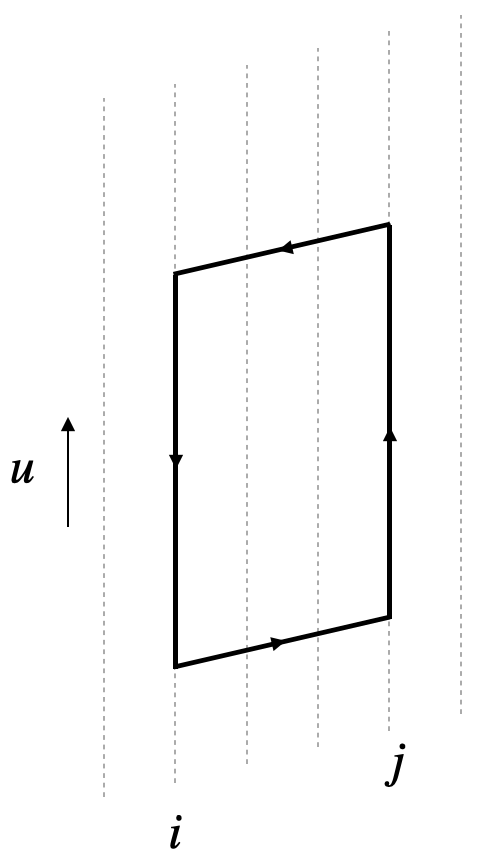}
\caption{A Wilson loop describing the evolution of a charge-anticharge pair at sites $j$ and $i$ respectively.\label{fig:temporal1}}
\end{center}
\end{figure}

It follows that $\langle \mathcal{W}_{\mathcal{C}}\rangle = 0$ but the two-point function of Wilson loops is akin to the two-point function~\eqref{E:compactVertex} of $e^{i q \varphi}$ for the compact scalar,
\beq
	\langle \mathcal{W}_{\mathcal{C}}^{q_1}(u) \mathcal{W}_{\mathcal{C}'}^{(q_2)}(0)\rangle = e^{-i E_{\mathcal{C}}^{(q_2)} u} \left( \delta_{q_1,-q_2} \delta_{\mathcal{C},\mathcal{C}'} + \delta_{q_1,q_2}\delta_{\mathcal{C},-\mathcal{C}'}\right)\,.
\eeq
However, Wilson loops partially extended in the temporal direction can have non-vanishing expectation values. As an example consider a spacetime curve $\mathcal{C}$ which is a union of four segments:
\begin{enumerate}
	\item A spatial curve $\mathcal{C}_s$ connecting site $i$ to site $j$ at time $0$.
	\item A temporal curve $\mathcal{C}_j$ extending up in time at site $j$ from time $0$ to time $u$.
	\item A spatial curve $\mathcal{C}_{s'}$ connecting site $j$ to site $i$ at time $u$.
	\item A reversed temporal curve $\mathcal{C}_i$ at site $i$ from time $u$ to time $0$.
\end{enumerate}
See Fig.~\ref{fig:temporal1}.  The Wilson loop corresponding to transporting a charge $q$ around this loop has the expectation value identical to the two-point function of Wilson lines in the unconstrained theory, one line inserted along $\mathcal{C}_s$ at time $0$ and the other along $\mathcal{C}_{s'}$ at time $u$, so that
\beq
	\langle \mathcal{W}_{\mathcal{C}}^{(q)}\rangle = e^{- i E_{\mathcal{C}_s}^{(q)}u} \delta_{\mathcal{C}_s,\mathcal{C}_{s'}}\,,
\eeq
i.e.\ the spatial curves have to be identical. With $g_{\rm eff} = a^{d/2} g$ the Euclidean version goes as $\sim e^{- \frac{a g^2q^2}{2} L_{\mathcal{C}_s} T}$ with $T$ the extent in Euclidean time and $L_{\mathcal{C}_s}$ the physical length of the spatial curve. Before taking the continuum limit we see that there is a sense in which Wilson loops have an area law, although the effective string tension $\frac{a g^2q^2}{2}$ vanishes in the $a \to 0$ limit.

Our analysis here resembles in some ways the strong coupling limit of nonlinear sigma models~\cite{PhysRevD.19.3091} and relativistic gauge theories~\cite{PhysRevD.11.395}. In the latter the electric field is also conserved link-by-link and its vacuum correlation functions vanish. However the resemblance ends there, and the Carrollian and strong coupling limits differ. After all, the former breaks Lorentz symmetry while the latter preserves it. But, more importantly, $H=0$ in the string coupling limit while the Hamiltonian of the Carrollian theory remains non-trivial.

So far we have discussed Carrollian electromagnetism in $A_u = 0$ gauge. When we study a Carrollian version of scalar QED in Subsection~\ref{S:electricQED} it will be convenient to switch to a Coulomb gauge in which the product of exponentiated link variables $e^{i\varphi}$ entering each site equals unity. In the continuum this gauge choice is equivalent to the usual condition $\nabla \cdot A = 0$. Coulomb gauge has the property that the coupling between $A_u$ and the link variables vanishes: that coupling goes as the $u$-derivative of the gauge-fixing condition. As a result we can analyze the behavior of $A_u$ and the link variables separately. This property will be important, since ``electric matter'' only couples to $A_u$, so that the loop corrections to correlation functions of gauge-invariant matter operators are insensitive to the link variables and so to all of the subtleties studied in this Subsection.

\section{``Magnetic theories''}
\label{S:magnetic}

\subsection{Setup and Hamiltonian framework}

Now we switch gears and consider so-called ``magnetic theories,'' which have actions of the form
\begin{align}
\label{E:magnetic1}
S_M = \int du\,d^d x \left(\chi \partial_uQ - \mathcal{H}(\chi, \nabla\chi,\hdots)\right)
\end{align}
where the Hamiltonian density $\mathcal{H}(\chi, \nabla\chi,\hdots)$ only depends on $\chi$ and its spatial derivatives.  Such theories are Carroll-invariant and in fact possess a larger invariance under ``supertranslations'' $u \to u + f(x)$. With flat space holography in mind, it will sometimes be prudent to consider magnetic theories on the manifold $\mathbb{R} \times \mathbb{S}^d$, in which case the derivatives of $\chi$ arising in the second term become covariant spatial derivatives on the sphere.

A characteristic feature of~\eqref{E:magnetic1} is that $\chi$ has no dynamics, and as such is effectively frozen on a time slice. To see this integrate the $\chi \partial_uQ$ portion of the action by parts, giving
\begin{align}
\int du\,d^d x \left(- (\partial_u\chi) Q - \mathcal{H}(\chi, \nabla\chi,\hdots)\right)\,,
\end{align}
so $Q$ is a Lagrange multiplier imposing $\partial_u\chi = 0$, i.e.~$\chi = \chi(x)$ is independent of time.  Thus in the path integral, integration over $Q$ constrains us to the submanifold of time-independent configurations $\chi(x)$.  When the magnetic theory lives on $\mathbb{R} \times \mathbb{S}^d$ the constraint localizes the $\chi$ fields to live on the sphere $\mathbb{S}^d$.\footnote{From a Carroll group perspective the absence of time-dependence can be seen as follows.  The energy is a central element of the Carroll group. On the zero energy surfaces one obtains two further invariants which are the square of momentum and center of mass~\cite{Duval:2014uoa}. When one restricts to zero center of mass, Carroll acts as the Euclidean group.}

Now let us compute the basic quantum mechanical properties of the magnetic theory~\eqref{E:magnetic1}, inspired by the analysis in~\cite{3dflat}.  By analogy with single particle quantum mechanics, it is convenient to label a basis of states in the theory by fixed-time field configurations $\chi(x)$.  The basis is then $\{|\chi(x)\rangle\}$ ranging over all $\chi(x)$.  To compute the inner product of such states $\langle \chi_2(x) | \chi_1(x)\rangle$, we can compute the transition amplitude
\begin{align}
\lim_{\delta u \to 0}\langle \chi_2(x) |\,e^{- i \widehat{H} \delta u}\,|\chi_1(x)\rangle = \langle \chi_2(x) | \chi_1(x)\rangle\,.
\end{align}
To compute the left-hand side using the path integral we expand $\chi(u,x)$ around a straight-line trajectory as
\begin{align}
	\chi(u,x) = \chi_1(x) + \frac{u}{\delta u} (\chi_2(x) - \chi_1(x)) + \delta \chi(u,x)\,,
\end{align}
where $0 \leq u \leq \delta u$ and $\delta \chi(0,x) = \delta \chi(\delta u,x) = 0$. Above, $\chi(u,x)$ interpolates between $\chi_1(x)$ and $\chi_2(x)$ in a time interval $\delta u$.  The action is then
\begin{align}
\int_0^{\delta u} \! du\int d^d x \left(-(\partial_u\chi)Q - \mathcal{H}(\chi, \nabla \chi,...)\right) = -\int d^d x\,Q(0,x)\,(\chi_2(x) - \chi_1(x)) \, + O(\delta u)
\end{align}
at small $\delta u$.  Rewriting $Q(0,x)$ as $Q(x)$, the inner product $\langle \chi_2(x) | \chi_1(x)\rangle$ is given by
\begin{align}
\langle \chi_2(x)| \chi_1(x)\rangle = \int [dQ(x)]\,e^{-i \int d^d x \, Q(x) \left(\chi_2(x) - \chi_1(x)\right)} =: \delta[\chi_2(x) - \chi_1(x)]\,,
\end{align}
and the states $|\chi(x)\rangle$ have the appropriate Dirac delta functional inner product. Moreover, the $|\chi(x)\rangle$'s are eigenstates of the Hamiltonian operator, 
\begin{align}
\widehat{H} |\chi(x)\rangle = E[\chi(x)]\,|\chi(x)\rangle
\end{align}
with $E[\chi(x)] := \int d^d x \, \mathcal{H}(\chi, \nabla \chi,...)$.  Indeed, in the Schr\"{o}dinger picture we have 
\begin{align}
[\widehat{Q}(x),\,\widehat{\chi}(y)] = i\,\delta^d(x-y)\,,\qquad [\widehat{H}(x),\,\widehat{\chi}(y)] = 0\,.
\end{align}

With our $\chi$-basis and inner product at hand, we can compute many quantities in the magnetic theory~\eqref{E:magnetic1}.  Below, we compute $S$-matrix elements, partition functions, and correlators.

\subsection{$S$-matrix and partition function}

We immediately obtain the $S$-matrix of~\eqref{E:magnetic1} in the $\chi$-basis,
\begin{align}
\label{E:SmatrixPbasis1}
\langle \chi_2(x)|\,e^{- i \widehat{H}\,u} |\chi_1(x)\rangle = e^{- i E[\chi_1]\,u}\, \delta[\chi_2(x) - \chi_1(x)]\,.
\end{align}
Because $\chi$ is locally conserved, we note the theory~\eqref{E:magnetic1} has a continuous spectrum of energy eigenstates, which in turn implies an IR-divergent partition function
\begin{align}
\text{tr}(e^{- \beta \widehat{H}}) = \int [d\chi(x)]\, \langle \chi(x)|\,e^{- \beta \widehat{H}}\,|\chi(x)\rangle = \delta[0]\left( \int [d\chi(x)]\,e^{- \beta\, E[\chi]}\right).
\end{align}
This divergence has been observed in prior works (e.g.~\cite{deBoer:2023fnj}). We observe that it is an inevitable consequence of a continuous spectrum, just as one finds for a free quantum mechanical particle. Upon stripping off the prefactor $\delta[0]$ we can obtain an IR-finite notion of a partition function, $\mathcal{Z}(\beta) = \int [d\chi(x)]e^{-\beta E[\chi]}$, which we recognize as a statistical field theory partition function in $d$-dimensions. In this way we see that the $d+1$-dimensional magnetic theory can be interpreted as a $d$-dimensional statistical system.

As a concrete example, let us consider the magnetic theory
\begin{align}
S_M = \int du \,d^2 \Omega \left( - (\partial_u \chi)Q - \frac{1}{2}\left(M^2 \chi^2 + \frac{1}{R^2} |\nabla \chi |^2\right)\right)
\end{align}
on $\mathbb{R} \times \mathbb{S}^2$ where the $\mathbb{S}^2$ has radius $R$, and where $\nabla$ denotes the covariant derivative on $\mathbb{S}^2$.  Versions of this theory were studied in~\cite{Henneaux:2021yzg, deBoer:2021jej}.  Expanding $\chi$ on a fixed time slice in spherical harmonics as
\begin{align}
\chi(\Omega) = \sum_{\ell \geq |m|\geq 0}  c_{\ell,m}\, Y_{\ell m}(\Omega)
\end{align}
where $c_{\ell, m}^* = (-1)^m c_{\ell, -m}$ so that $\chi(\Omega)$ is real-valued, and taking the inner product on the space of field configurations to be
\beq
	(\delta \chi_1,\delta \chi_2) = \frac{1}{2\pi}\int du \,d^2\Omega \,\delta \chi_1 \delta\chi_2\,,
\eeq
we find
\begin{align}
\langle \chi'(\Omega) | \chi(\Omega)\rangle = \prod_{\ell\geq |m|\geq 0} \delta(c_{\ell,m}'-c_{\ell,m})\,,
\end{align}
as well as
\begin{align}
E[\chi(\Omega)] = \frac{1}{2}\sum_{\ell \geq |m|\geq 0}  \left(M^2 + \frac{\ell(\ell+1)}{R^2}\right) |c_{\ell,m}|^2\,.
\end{align}
To emphasize the dependence of $E[\chi(\Omega)]$ on the $c_{\ell,m}$, we will also write the energy as $E[\{c_{\ell,m}\}]$. In the $\chi$-basis, the $S$-matrix is
\begin{align}
\langle \chi'(\Omega)|\,e^{- i \widehat{H}\,u} |\chi(\Omega)\rangle = e^{- i E[\{c_{\ell,m}\}]\,u} \prod_{\ell\geq |m|\geq 0}\delta(c_{\ell,m}' - c_{\ell,m})\,,
\end{align}
and the finite version of the thermal partition function is that of a statistical field theory of $\chi$ on a sphere,
\begin{align}
\mathcal{Z}=  \prod_{\ell \geq 0} \left( \beta \left( M^2 + \frac{\ell(\ell+1)}{R^2} \right)\right)^{-\left(\ell + \frac{1}{2}\right)} \,.
\end{align}
The infinite product may be regularized in any number of ways and the answer is unique up to a choice of local counterterms. The second derivative of its logarithm with respect to $M^2$ is unambiguous and reads
\begin{align}
\frac{\partial^2\log \mathcal{Z}}{\partial (M^2)^2}= - \frac{R^4}{2}\, \frac{\psi^{(1)}(\Delta_+) - \psi^{(1)}(\Delta_-)}{\Delta_+ - \Delta_-}
\end{align}
where $\Delta_\pm := \frac{1}{2}\left(1 \pm \sqrt{1 - 4 M^2 R^2}\right)$ and $\psi^{(1)}$ is the derivative of the Euler digamma function.

\subsection{Correlation functions}
\label{sec:magneticcorrelation}

Computing correlators of $\chi$ alone is rather straightforward. Suppose that we wish to compute thermal correlation functions at inverse temperate $\beta$. Considering the thermal density matrix $\hat{\rho}_\beta = \frac{1}{Z(\beta)}\,e^{- \beta \widehat{H}}$, we simply have
\begin{align}
\text{tr}\!\left(\hat{\rho}_\beta\,\widehat{\chi}(u_1,x_1) \cdots \widehat{\chi}(u_{n},x_n)\right) = \frac{1}{Z(\beta)}\int [d\chi(x)]\,e^{- \beta\,E[\chi]}\,\chi(x_1) \cdots \chi(x_n)\,,
\end{align}
i.e.\ correlations of $\chi$ reduce to those of a $d$-dimensional statistical system. In particular they are independent of time $u$ and therefore ``magnetic.'' The correlators of $Q$ are more interesting. On the one hand, $Q$ is the canonical conjugate to $\chi$, so to produce $Q$ insertions we can simply differentiate amplitudes with respect to  $\chi$. However, as when computing correlation functions of the position for a free quantum mechanical particle, the $Q$ correlators tend to be IR-divergent on account of the zero mode of $Q$. To avoid that problem we instead consider correlators of $\frac{dQ}{du}$.   First we observe that the Heisenberg equations of motion give us
\begin{align}
\frac{d\widehat{Q}}{du} = i\,[\widehat{H}, \widehat{Q}] = \frac{\delta E[\widehat{\chi}]}{\delta \widehat{\chi}(x)}\,,
\end{align}
which we can exactly solve by
\begin{align}
\widehat{Q}(u,x) =  \frac{\delta E[\widehat{\chi}]}{\delta \widehat{\chi}(x)}\,u + \widehat{Q}(0,x)\,.
\end{align}
But now, apart from contact terms, we have reduced the problem of computing correlators of $\frac{dQ}{du}$ to that of the composite $\frac{\delta E}{\delta \chi}$ built from $\chi$. In particular $\frac{dQ}{du}$ has magnetic correlations. This implies e.g.
\begin{align}
\text{tr}\!\left(\hat{\rho}_\beta\,\frac{d\widehat{Q}(u,x)}{du}\right) &= \int [d\chi(x)]\, \frac{\delta E[\chi]}{\delta \chi}\,e^{-\beta\,E[\chi]} = 0\,.
\end{align}

\section{Putting the pieces together}
\label{S:electromagnetic}

At the beginning of this manuscript we observed in~\eqref{E:twoPoint} that unbroken supertranslation and ordinary translational invariance implies that two-point functions decompose into a sum of two parts that can be called electric and magnetic. In a generic theory one might imagine that there is a choice of operator basis such that the matrix of two-point functions is either purely electric or purely magnetic. However, higher point functions are not constrained in the same way, and so there may be mixing of the two sectors. Even so, supertranslations do impose strong constraints, namely that correlation function for each pair of operators either depends on the operators' spatial separation with no time dependence, or on the operators' temporal separation with a spatial $\delta$ function. Concretely three-point functions are fixed to have the form~\cite{Bagchi:2023cen}
\begin{align}
\begin{split}
 \label{eq:3pt_mixed}
    \langle\mathcal{O}_1(u_1,x_1)  \mathcal{O}_2(u_2,x_2)  \mathcal{O}_3(u_3,x_3) \rangle &= F_{\rm MMM}(x_{12},x_{23}) + F_{\rm EEE}(u_{12},u_{23}) \delta(x_{12})\delta(x_{23}) 
    \\
    &~~~~~ \quad \quad \quad + \left(F_{\rm EEM}(u_{12},x_{23}) \delta(x_{12}) + (\text{permutations}) \right) \,.
\end{split}
\end{align}
We emphasize that even if $\mathcal{O}_1, \mathcal{O}_2$ are pure electric and $\mathcal{O}_3$ is pure magnetic at the level of two-point functions, then $F_{\rm EEM}$ need not vanish.

The above structure describes spatial interactions between electric operators for high-point correlation functions. For example, the four-point function of purely electric operators can contain a piece the form
\begin{equation} 
\label{eq:4pt_mixed}
    \langle\mathcal{O}_1(u_1,x_1) \mathcal{O}_2(u_2,x_2) \mathcal{O}_3(u_3,x_3) \mathcal{O}_4(u_4,x_4) \rangle \supset \delta(x_{12}) \delta(x_{34}) F_{1234}(u_{12},u_{34},x_{23}) + (\text{permutations}) \,.
\end{equation}

In the remainder of the section we explore two examples of coupled Carrollian field theories that generate such correlations. Specifically, we study a scalar Yukawa theory coupling an electric scalar to a magnetic one, and a Carrollian version of scalar QED. One difficulty in studying these examples is that new UV divergences are generated which must be tamed in a consistent manner. In particular, the short-distance behavior of the magnetic theory appears in loops, which creates a form of UV/IR mixing. Fortunately the same strategies we took in Sections~\ref{S:electric2} and~\ref{S:EM} allow us to make sense of these theories, namely to regulate them on a spatial lattice and take a suitable continuum limit.

\subsection{Scalar-Yukawa theory}

\subsubsection{Finite temperature and UV/IR mixing}

We begin by coupling a free electric scalar to a Gaussian magnetic theory through a Yukawa interaction. To ameliorate UV and IR divergences and ultimately make contact with a Hamiltonian formulation, we work at finite temperature $T = 1/\beta$ and on a hypercubic space lattice $\Gamma$ with lattice scale $a$.  We will ultimately consider the regime where $a \to 0$, and where $\beta$ is either finite or taken to infinity. As in our discussion at the end of Subsection~\ref{S:electric2} we can reliably compute using perturbation theory through a suitable continuum prescription. In the imaginary time $\tau = i u$ formalism the Lagrangian is
\begin{align}
\label{E:YukawaLag1}
    \mathcal{L} = \frac{1}{2} (\partial_{\tau}\phi)^2 + \frac{m^2}{2} \phi^2 + i \chi \partial_{\tau}Q + \frac{1}{2}|\nabla \chi|^2 +\frac{M^2}{2} \chi^2 + \frac{\lambda a^{d/2}}{2} \chi \phi^2\,,
\end{align}
where we have put a factor of $a^{d/2}$ in front of the Yukawa coupling $\lambda$. This unusual scaling of $\lambda$ can be regarded as a multiplicative renormalization which renders the theory UV finite. While the renormalization of couplings is usually additive in relativistic field theories, a multiplicative renormalization of couplings arises in field theories with dipole or subsystem symmetry~\cite{dipoleRG}.

There is one other piece of data we need to define correlators in the Yukawa theory.  Notice that the tree level $\chi \chi$ correlator is $\langle \chi(x,\tau) \chi(y,\tau')\rangle_0 = \frac{1}{\beta}\,D(x-y)$ for $D(x-y) := \int\frac{d^d k}{(2\pi)^d} \frac{e^{i k \cdot (x-y)}}{k^2 + M^2}$, which in the continuum diverges at $x = y$.  In our lattice regularization, $D(0)$ is really $ \kappa /a^d$ with $\kappa$ a non-universal coefficient depending on the lattice regularization.  The coefficient $\kappa$ will survive in the $a \to 0$ limit of our correlators, and can be viewed as a piece of UV data which is salient in the IR, and hence manifests UV/IR mixing.  We also note that if we sought to analyze~\eqref{E:YukawaLag1} non-perturbatively, it would be useful to replace the terms $\frac{M^2}{2}\,\chi^2 + \frac{a^{d/2}\lambda}{2}\,\chi \phi^2$ by, for instance, $\frac{1}{2}(M\chi + \frac{a^{d/2}\lambda}{2M}\,\phi^2)^2 =  \frac{M^2}{2} \chi^2 + \frac{a^{d/2}\lambda}{2}\chi \phi^2 + \frac{a^d \lambda^2}{8M^2} \phi^4$, giving a bounded potential. But we will not pursue this modification here.

This theory has the nice feature that we can compute directly in the continuum limit using a prescription similar to that employed for pure electric theories at the end of Subsection~\ref{S:electric2}. There is a single saddle point and fluctuations around it are encoded by lattice perturbation theory. With the scaling limit of couplings indicated in~\eqref{E:YukawaLag1}, lattice perturbation theory has an $a\to 0$ limit in terms of continuum diagrams. In this continuum prescription interaction vertices are equipped with powers of the lattice spacing as indicated by~\eqref{E:YukawaLag1}; in various places we require the lattice identity $\delta(0) = a^{-d}$; and, as mentioned above, we also require the short distance limit of the $\chi$ propagator $D(0) = \kappa a^{-d}$. With these powers of $a$ explicitly included, connected $n$-point diagrams of fundamental fields scale as $a^{\frac{d(n-2)}{2}}$ just as in pure electric theories.

It will be useful to define
\begin{align}
\label{E:Kbeta}
K_\beta(\tau-\tau') := \frac{1}{\beta} \sum_n e^{i \omega_n (\tau-\tau')} \frac{1}{\omega_n^2 + m^2} = \frac{1}{2m} \frac{\cosh\!\left((\frac{\beta}{2} - |\tau - \tau'|) m\right)}{\sinh(\frac{\beta m}{2})}\,,
\end{align}
where we will henceforth assume that $|\tau - \tau'| \leq \beta$. With this in hand we begin with the loop corrected $\phi$-propagator, which has the diagrammatic expansion
\begin{center}
\includegraphics[scale=.5]{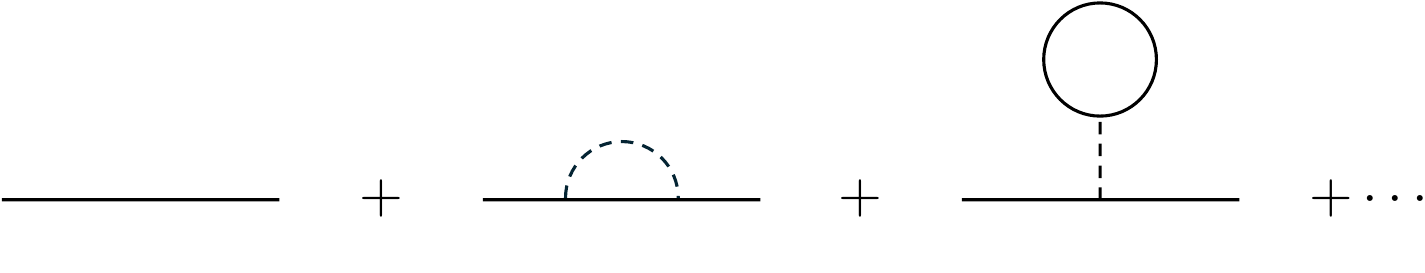}
\end{center}
Here the solid lines correspond to $\phi$ propagators and the dashed lines corresponds to $\chi$ propagators.  In the $a \to 0$ limit, the above diagrams yield
\begin{align}
        \langle \phi(\tau,x)\phi(\tau',y) \rangle_{\rm conn} &= \delta(x - y) \Bigg[K_\beta(\tau - \tau') + \frac{\lambda^2 \kappa}{2\beta}\frac{\partial}{\partial m^2}\frac{\partial}{\partial m^2}K_\beta(\tau - \tau') 
        \\
        \nonumber
        &\qquad \qquad \qquad \qquad \qquad \qquad - \frac{\lambda^2}{2 M^2}\,K_\beta(0) \left(\frac{\partial}{\partial m^2} K_\beta(\tau - \tau')\right) +O(\lambda^4)  \Bigg]\,.
\end{align}
We observe that the first loop correction depends on $\kappa/\beta$ which reflects both a UV sensitivity in $\kappa$ and an IR sensitivity in $\beta$. The second correction depends on the short time limit of $K_{\beta}$, $K_{\beta}(0) = \frac{1}{2m}\coth\!\left(\frac{\beta m}{2}\right)$, which also contributes to the one point function $\langle \chi \rangle$.  At one-loop the tadpole comes from
\begin{center}
\vspace{-.05in}
\includegraphics[scale=.5]{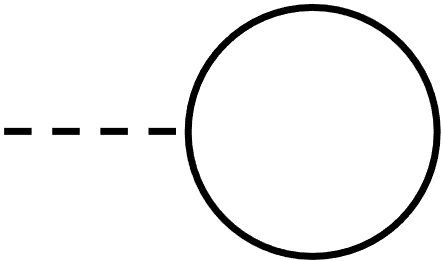}
\vspace{-.05in}
\end{center}
which evaluates to
\begin{align}
 \langle \chi\rangle&= - \frac{\lambda}{2 M^2}\,K_\beta(0)a^{-d/2} \,+ O(\lambda^3)\,.
\end{align}
As for one-point functions of pure electric theories (recall that connected $n$-point functions there go as $a^{\frac{d(n-2)}{2}}$), we see that $\langle \chi \rangle$ diverges in the continuum limit with the expected power $\sim a^{-d/2}$.

Next we turn to the loop-corrected $\chi$-propagator, corresponding to the diagrams
\begin{center}
\includegraphics[scale=.5]{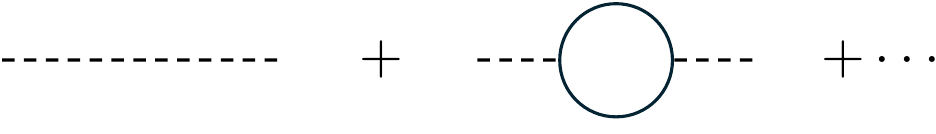}
\end{center}
which gives
\begin{align}
  \langle \chi(\tau,x)\chi(\tau',y) \rangle_{\rm conn} &= \frac{1}{\beta}\,D(x - y)+ \frac{\lambda^2}{2\beta} \left(\frac{\partial}{\partial m^2}\,K_\beta(0)\right) \! \left(\frac{\partial}{\partial M^2}\,D(x-y)\right) \,+ O(\lambda^4)\,.
\end{align}

It is also natural to consider correlators of $\partial_{\tau}Q$.  The reason we consider the derivative instead of $Q$ is to obviate the need to regulate the temporal zero mode. For $\langle \partial_{\tau}Q \rangle$ we have the diagram
\begin{center}
\includegraphics[scale=.5]{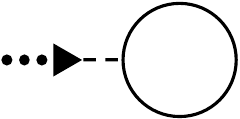}
\end{center}
where the dotted lines connected to an arrow connected to dashed lines correspond to a $(\partial_{\tau}Q)\chi$ propagator.  The one-loop correction vanishes, 
\begin{align}
\langle \partial_\tau Q\rangle = 0 +\,O(\lambda^2)\,.
\end{align}
Similarly the one-loop corrected version of $\langle \partial_{\tau}Q \partial_{\tau}Q\rangle$ is
\begin{center}
\includegraphics[scale=.5]{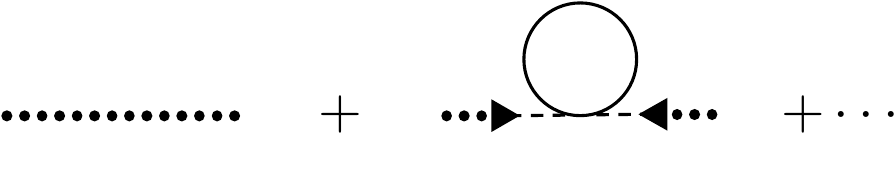}
\end{center}
where the dotted lines correspond to a $\partial_{\tau}Q \partial_{\tau}Q$ propagator, while $\langle (\partial_{\tau} Q) \chi\rangle$ is
\begin{center}
\includegraphics[scale=.5]{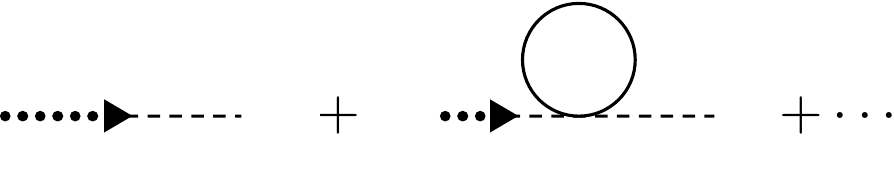}
\end{center}
These give
\begin{align}
\langle \partial_{\tau}Q(\tau,x)\partial_{\tau'}Q(\tau',y)\rangle &= (- \nabla^2 + M^2) \delta(x-y)\left(\delta(\tau - \tau') - \frac{1}{\beta}\right) \\
& \qquad \qquad \qquad \qquad - \frac{\lambda^2}{2}\,\delta(x-y)\left((K_\beta(\tau - \tau'))^2 + \frac{1}{\beta} \frac{\partial}{\partial m^2} K_\beta(0)\right) + \,O(\lambda^4)\,,
 \nonumber 
 \\
\langle \partial_{\tau}Q(\tau,x) \chi(\tau',y)\rangle &= i \, \delta(x-y)\left(\delta(\tau - \tau') - \frac{1}{\beta}\right) + \,O(\lambda^4)\,.
\end{align}

We can also take the zero temperature limit $\beta \to \infty$ (for $m\beta \gg 1$) of all of the correlators above without introducing new divergences:
\begin{align} 
\begin{split}\label{eq:yukawazeroT}
        \langle \phi(\tau,x)\phi(\tau',y) \rangle &= \frac{e^{-m|\tau-\tau'|}}{2m} \left(1 + \frac{\lambda^2}{8 M^2 m^3}(1 + m\,|\tau - \tau'|)\,+O(\lambda^4)\right)\delta(x - y) \,,
        \\
        \langle \chi\rangle&= - \frac{\lambda}{4 a^{d/2} M^2 m}\,+ O(\lambda^3) \,, 
        \\ 
        \langle \chi(\tau,x)\chi(\tau',y) \rangle &= \langle \chi \rangle^2 \,+ O(\lambda^4)\,,
        \\
        \langle \partial_{\tau}Q\rangle &= 0 +\,O(\lambda^2)\,, 
        \\ 
        \langle \partial_{\tau}Q(\tau,x)\partial_{\tau'}Q(\tau',y)\rangle &= (-\nabla^2 + M^2)\delta(x-y)\,\delta(\tau - \tau') - \lambda^2\,\delta(x-y)\,\frac{e^{- 2 m |\tau - \tau'|}}{8 m^2} +\,O(\lambda^4)\,, 
        \\
        \langle \partial_{\tau}Q(\tau,x)\chi(\tau',y)\rangle &= i\,\delta(x-y)\,\delta(\tau-\tau') + \,O(\lambda^4)\,.
\end{split}
\end{align}
Later we will provide a Hamiltonian formulation of the scalar Yukawa theory, and give an example computation of a $\beta \to \infty$ correlator.

While so far we have considered one point and two point functions, our techniques readily generalize to higher point functions as well as to correlation functions of composite operators. For example, at finite $\beta$ and at tree level, the leading contribution to the connected three point function $\langle \phi \phi \chi\rangle$ comes from
\begin{center}
\includegraphics[scale=.5]{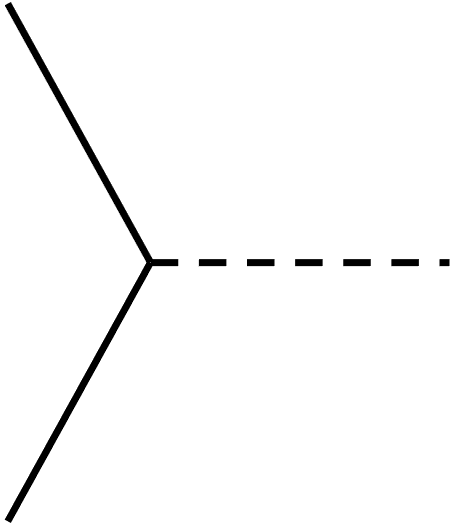}
\end{center}
which gives
\begin{align}
\langle \phi(\tau_1,x_1) \phi(\tau_2,x_2) \chi(\tau_3,x_3)\rangle_{\text{conn}} = \frac{\lambda a^{d/2}}{\beta}\,D(x_1 - x_3)\,\delta(x_1 - x_2)\,\frac{\partial}{\partial m^2}\,K_\beta(\tau_1 - \tau_2) +\,O(\lambda^3)\,.
\end{align}
Similarly, at finite $\beta$ and at tree level, the connected four point function $\langle \phi \phi \phi \phi \rangle$ corresponds to the diagrams
\begin{center}
\includegraphics[scale=.5]{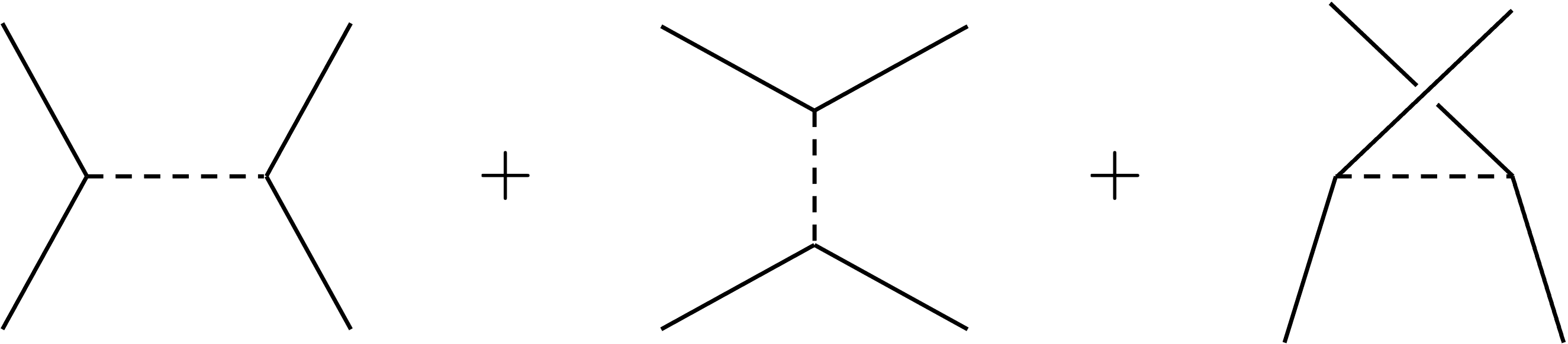}
\end{center}
and reads
\begin{align}
&\langle \phi(\tau_1,x_1) \phi(\tau_2,x_2) \phi(\tau_3,x_3) \phi(\tau_4,x_4)\rangle_{\text{conn}} 
\\
& \qquad \qquad = \frac{\lambda^2 a^d}{\beta}\,D(x_1-x_3)\,\delta^d(x_1-x_2)\,\delta^d(x_3-x_4)\,\left(\frac{\partial}{\partial m^2}\,K_\beta(\tau_1 - \tau_2)\right)\!\left(\frac{\partial}{\partial m^2}\,K_\beta(\tau_3 - \tau_4)\right) \nonumber 
\\
& \qquad \qquad \qquad \qquad \qquad \qquad \qquad \qquad \qquad \qquad \qquad \qquad  \qquad \qquad \qquad +\,(\text{permutations})\,+\,O(\lambda^4)\,. \nonumber
\end{align}
Notice that as $a \to 0$ the three- and four-point functions vanish as $a^{d/2}$ and $a^d$ respectively. This is the same scaling we observed in pure electric theories, and it is easy to see that this scaling is maintained at loop-level in the coupled theory. Evidently the scalar Yukawa theory is also a model of Gaussian correlations. We will see a similar phenomena later when we study Carrollian scalar QED.

\subsubsection{Hamiltonian formalism}

It is enlightening to compare the perturbative diagrammatic expansion given above to the perturbation theory in the Hamiltonian formalism. The Hamiltonian of our scalar Yukawa theory is
\begin{equation}
\label{E:YukawaHam1}
H =  \sum_{i\in\Gamma}\left\{ \frac{\pi_i^2}{2a^d} + a^d\left( \frac{1}{2} m^2 \phi_i^2 + \frac{1}{2} |\nabla \chi_i|^2 + \frac{M^2}{2} \chi^2_i + \frac{\lambda a^{d/2}}{2} \chi_i\phi_i^2\right) \right\}\,,
\end{equation}
where $\pi$ is the canonical conjugate field of $\phi$ and $|\nabla \chi_i|^2 $ is a shorthand for the discretized derivative squared. As the Hamiltonian is independent of $Q$, the canonical conjugate field of $\chi$ is a constant of motion.  We can reproduce our finite-temperature correlators in the Hamiltonian formalism using the Gibbs state
\begin{align}
\hat{\rho}_\beta = \frac{1}{Z(\beta)}\,e^{- \beta \widehat{H}}\,,
\end{align}
where $\widehat{H}$ is the operator version of~\eqref{E:YukawaHam1}.  To see how this works we compute the zero-temperature, continuum limit of $\langle \phi \phi\rangle_{\rm conn}$.

The basic idea is that we can exploit the conservation of $\chi$ to regard this model as an effective quantum mechanics for the $\phi_i$. In a state of definite $\chi$ we have a collection of oscillators $\phi_i$ with a position-dependent frequency $(\omega_{\rm eff}^{\chi})^2 = m^2 + \lambda a^{d/2}\chi$. The field $\chi$ acts as a statistical background for $\phi$ and is thermally distributed. Correlation functions can then be obtained through a combination of quantum mechanical perturbation theory for the oscillators, and a residual sum over the thermal distribution for $\chi$.

It will be useful to develop some notation for the $\beta \to \infty$ limit of $\hat{\rho}_\beta$.   To this end let $|0^{\chi}_i\rangle$ be the ground state of the $i$th oscillator at fixed $\chi$, $(\mathcal{E}_0^\chi)_i= \frac{1}{2}((\omega_{\text{eff}}^\chi)_i  - m)$ the (shifted) ground state energy, and $|0^\chi\rangle = \bigotimes_{i\in\Gamma} \left(|0^\chi_i\rangle \otimes |\chi_i\rangle\right)$ the simultaneous ground state at fixed $\chi$. We take the $|0_i^{\chi}\rangle$'s to be normalized such that $\langle 0_i^{\chi} |0_j^{\chi}\rangle = \frac{\delta_{ij}}{a^d}$.  Then we have
\begin{align}
\label{E:rhobetatoinf1}
\hat{\rho}_{\beta \to \infty} \approx \frac{1}{Z(\beta)} \int \prod_{i\in\Gamma}d\chi_i\,e^{-\beta a^d\sum_{i\in\Gamma}\left\{ \frac{1}{2} |\nabla \chi_i|^2 + \frac{M^2}{2} \chi^2_i\right\} - \beta\sum_{i\in\Gamma}(\mathcal{E}_0^\chi)_i} |0^\chi\rangle\langle 0^\chi|\,.
\end{align}
Later on we will alternatively write the above as $\lim_{\beta \to \infty} \int [d\chi]\,P_\beta[\chi]\,|0^\chi\rangle \langle 0^\chi|$ or $\lim_{\beta \to \infty} \mathbb{E}_{\chi \sim P_\beta}\!\left[|0^\chi\rangle \langle 0^\chi|\right]$.

Using~\eqref{E:rhobetatoinf1} and defining the effective Hamiltonian at fixed $\chi$, namely $\widehat{H}_\phi[\chi] := \sum_{i\in\Gamma}\left\{\frac{\hat{\pi}_i^2}{2a^d} + \frac{a^d}{2}(\omega_{\text{eff}}^\chi)_i^2 \hat{\phi}_i^2\right\}$, we obtain
\begin{align}
       &\text{tr}\big(\hat{\rho}_{\beta \to \infty}\hat{\phi}_i(\tau) \hat{\phi}_j(\tau')\big)   \nonumber 
       \\
        &\quad =\frac{1}{Z(\beta)} \int\prod_i d\chi_i\,e^{-\beta a^d\sum_{i\in\Gamma}\left\{ \frac{1}{2} |\nabla \chi_i(x)|^2 + \frac{M^2}{2} \chi^2_i\right\} - \beta\sum_{i\in\Gamma} (\mathcal{E}_0^\chi)_i} \nonumber 
        \\
        & \qquad \qquad \qquad \qquad \qquad \times \left\langle 0^{\chi}\right| e^{\widehat{H}_\phi[\chi] \,\tau} \hat{\phi}(x,0) e^{- \widehat{H}_\phi[\chi] \,(\tau - \tau')}  \hat{\phi}(y,0) e^{- \widehat{H}_\phi[\chi] \,\tau'} \left|0^{\chi}\right\rangle \nonumber 
        \\
    &\quad =\frac{1}{Z(\beta)} \int \prod_kd\chi_k\,e^{-\beta  a^d\sum_{k\in\Gamma} \left\{ \frac{1}{2} |\nabla \chi_k|^2 + \frac{M^2}{2} \chi^2_k\right\} - \beta\sum_{k\in\Gamma}\,(\mathcal{E}_0^\chi)_k}\frac{1}{2|(\omega_{\text{eff}}^\chi)_i|}\,e^{-( \omega_{\text{eff}}^\chi)_i |\tau - \tau'|}\,\frac{\delta_{ij}}{a^d} \nonumber 
        \\
        &\quad = \frac{e^{-m|\tau - \tau'|}}{2m} \left(1 + \frac{\lambda^2}{8 M^2 m^3} (1 + m\,|\tau - \tau'|) + \,O(\lambda^4) \right)\frac{\delta_{ij}}{a^d}\,.
\end{align}
In going from the second-to-last line to the last line, we have expanded $\omega_{\text{eff}}^\chi(x)$'s to second order in $\lambda$, and $e^{- \beta\sum_{k\in\Gamma} (\mathcal{E}_0^{\chi})_k}$ to first order (the second order piece in this expansion gives rise to a vacuum bubble pertaining to the overall normalization). The end result has a smooth continuum limit,
\beq
	\text{tr}\!\left( \hat{\rho}_{\beta\to\infty} \hat{\phi}(\tau,x)\hat{\phi}(\tau',y)\right) = \frac{e^{-m|\tau-\tau'|}}{2m}\left( 1 + \frac{\lambda^2}{8M^2m^3}(1+m|\tau-\tau'|) + O(\lambda^4)\right)\delta(x-y)\,,
\eeq
which agrees with~\eqref{eq:yukawazeroT}.

\subsubsection{$Q$ correlators and a soft theorem}

At the end of Section \ref{sec:magneticcorrelation}, we observed that $Q$-insertions are equivalent to derivations with respect to $\chi$. Here we will check a related statement explicitly in perturbation theory in the scalar Yukawa theory as $\beta\to \infty$. The result can be interpreted as a soft theorem.

We begin by giving a general argument.  Let us define
\begin{align}
\begin{split}
H_\phi[\chi] &:= a^d \sum_{i \in \Gamma}\left\{\frac{1}{2}|\nabla \chi_i|^2 + \frac{1}{2} M^2 \chi_i^2\right\} + \sum_{i \in \Gamma} (\mathcal{E}_0^\chi[\phi])_i \\
&\sim \int d^d x \left\{\frac{1}{2}|\nabla \chi(x)|^2 + \frac{1}{2} M^2 \chi(x)^2 + \frac{1}{a^d}\,\mathcal{E}_0^\chi[\phi](x)\right\}\,.
\end{split}
\end{align}
(See above~\eqref{E:rhobetatoinf1} for the definition of $\mathcal{E}_0^\chi$.)  Akin to our analysis in Section~\ref{sec:magneticcorrelation}, we have the operator equation
\begin{align}
\partial_u \widehat{Q}(u,x) = \left.\frac{\delta H_\phi[\chi(x)]}{\delta \chi(x)}\right|_{\substack{\!\!\!\!\chi \to \widehat{\chi}(x) \\ \phi \to \widehat{\phi}(u,x)}}\,.
\end{align}
Analytically continuing $u \to - i \tau$ and compactifying $\tau$ on the thermal $\beta$-circle, we find
\begin{align}
i \,\partial_\tau \widehat{Q}(\tau,x) = \left.\frac{\delta H_\phi[\chi(x)]}{\delta \chi(x)}\right|_{\substack{\!\!\!\!\chi \to \widehat{\chi}(x) \\ \phi \to \widehat{\phi}(\tau,x)}}\,.
\end{align}
Next we take the Fourier transform from $\tau$ to $\omega$ on both sides and take the $\omega \to 0$ limit, giving us
\begin{align}
\lim_{\omega \to 0}\omega \,\widehat{Q}(\omega,x) = \int_0^\beta d\tau \left.\frac{\delta H_\phi[\chi(x)]}{\delta \chi(x)}\right|_{\substack{\!\!\!\!\chi \to \widehat{\chi}(x) \\ \phi \to \widehat{\phi}(\tau,x)}}\,.
\end{align}
Leveraging this identity, we find
\begin{align}
\begin{split}
&\lim_{\omega \to 0}\mathbb{E}_{\chi \sim P_\beta}\!\left[\langle 0^\chi| \,\omega\,\widehat{Q}(\omega,x)\,\mathcal{F}[\,\widehat{\phi}\,]\,|0^\chi\rangle\right] \\
=\,&  \frac{1}{Z(\beta)} \int [d\chi(x)]\,e^{- \beta H_\phi[\chi(x)]} \langle 0^\chi| \int_0^\beta d\tau \left.\frac{\delta H_\phi[\chi(x)]}{\delta \chi(x)}\right|_{\substack{\phi \to \widehat{\phi}(\tau,x)}}\,\mathcal{F}[\,\widehat{\phi}\,]\,|0^\chi\rangle\,.
\end{split}
\end{align}
Taking the large $\beta$ limit, since $\phi$ is massive it gets frozen in its ground state configuration $\phi = 0$, so that the second line above gives us
\begin{align}
\begin{split}
& \lim_{\beta \to \infty}\frac{1}{Z(\beta)} \int [d\chi(x)]\,e^{- \beta H_\phi[\chi(x)]} \langle 0^\chi| \beta \frac{\delta H_0[\chi(x)]}{\delta \chi(x)}\,\mathcal{F}[\,\widehat{\phi}\,]\,|0^\chi\rangle \\
=\,&  \lim_{\beta \to \infty}\frac{1}{Z(\beta)} \int [d\chi(x)] \left(-\frac{\delta}{\delta \chi(x)}\,e^{- \beta H_\phi[\chi(x)]} \right)\langle 0^\chi| \mathcal{F}[\,\widehat{\phi}\,]\,|0^\chi\rangle  \\
=\,&  \lim_{\beta \to \infty}\frac{1}{Z(\beta)} \int [d\chi(x)] \,e^{- \beta H_\phi[\chi(x)]} \frac{\delta}{\delta \chi(x)}\langle 0^\chi| \mathcal{F}[\,\widehat{\phi}\,]\,|0^\chi\rangle  \\
=\,&  \lim_{\beta \to \infty} \mathbb{E}_{\chi \sim P_\beta}\!\left[\frac{\delta}{\delta \chi(x)}\langle 0^\chi| \mathcal{F}[\,\widehat{\phi}\,]\,|0^\chi\rangle\right] \, ,
\end{split}
\end{align}
where in going from the second to third line we have integrated by parts in $\chi$.  Altogether, we find
\begin{align}
\label{E:altogether1}
\lim_{\omega \to 0} \lim_{\beta \to \infty}\mathbb{E}_{\chi \sim P_\beta}\!\left[\langle 0^\chi| \,\omega\,\widehat{Q}(\omega,x)\,\mathcal{F}[\,\widehat{\phi}\,]\,|0^\chi\rangle\right]  = \lim_{\beta \to \infty} \mathbb{E}_{\chi \sim P_\beta}\!\left[\frac{\delta}{\delta \chi(x)}\langle 0^\chi| \mathcal{F}[\,\widehat{\phi}\,]\,|0^\chi\rangle\right] \, ,
\end{align}
which is our desired soft theorem.

We now demonstrate the general formula~\eqref{E:altogether1} in a specific case.  The two-point function of $Q$ and $\chi$ as $\beta \to \infty$ from~\eqref{eq:yukawazeroT} is
\begin{equation}
  \label{eq:5}
\langle Q(\omega,x)\chi(\omega',y)\rangle=\frac{2\pi}{\omega}\,\delta(x-y)\delta(\omega+\omega')+O(\lambda^4) \, ,
\end{equation}
while $\langle Q\phi\phi\rangle$ to leading order in $\lambda$ is
\begin{align}
  \label{eq:7}
  \langle Q(\omega_3,x_3)&\phi(\tau_1,x_1)\phi(\tau_2,x_2)\rangle=-\frac{1}{\omega_3}\frac{\lambda}{4m^2}\delta(x_1-x_2)\delta(x_1-x_3)\int^\infty_{-\infty}d \tau e^{-m|\tau-\tau_1|}e^{-m|\tau-\tau_2|}e^{-i\omega_3\tau}\nonumber\\
  &=-\frac{\lambda}{4m^2}\delta(x_1-x_2)\delta(x_1-x_3)\frac{4 e^{-\frac{1}{2}|\tau_{12}|(2m+i\omega_3)}m\left(\omega_3\cos\!\left(\frac{|\tau_{12}|\omega_3}{2}\right)+2m \sin\!\left(\frac{|\tau_{12}|\omega_3}{2}\right)\right)}{4m^2\omega_3^2+\omega_3^4} \, .
\end{align}
In the limit $\omega_3\to 0$ the above becomes
\begin{align}
  \label{eq:8}
  \lim_{\omega_3\to 0}\langle Q(\omega_3,x_3)\phi(\tau_1,x_1)\phi(\tau_2,x_2)\rangle=-\frac{1}{\omega_3}\frac{\lambda}{2m^2}(1+m|\tau_{12}|)\delta(x_3-x_1)\langle \phi(\tau_1,x_1)\phi(\tau_2,x_2)\rangle+O(\omega_3^0)\,,
\end{align}
 proportional to the two-point function of $\phi$.  Now we also have
 \begin{equation}
   \label{eq:9}
\lim_{\beta \to \infty}\mathbb{E}_{\chi \sim P_\beta}\!\left[\langle 0^\chi|\widehat{\phi}(\tau_1,x_1)\widehat{\phi}(\tau_2,x_2) |0^{\chi}\rangle\right]=\frac{1}{2|\omega_{\text{eff}}^{\chi_0}(x_1)|}e^{-\omega_{\text{eff}}^{\chi_0}(x_1)|\tau_1-\tau_2|} \delta(x_1-x_2)\,,
 \end{equation}
 where $\chi_0=-\frac{\lambda}{4a^{d/2} M^2m} +O(\lambda^3)$ is the $\chi$-vacuum corresponding to the 1-loop vacuum expectation value of $\chi$, and relatedly
\begin{align}
  \label{eq:10}
 &\lim_{\beta \to \infty}\mathbb{E}_{\chi \sim P_\beta}\!\!\left[\frac{\delta}{\delta \chi(x)}\langle 0^\chi|\widehat{\phi}(\tau_1,x_1)\widehat{\phi}(\tau_2,x_2) |0^\chi\rangle\right] \\
 & \qquad \qquad \qquad \qquad \qquad \qquad \qquad =-\frac{\lambda(1+\omega_{\text{eff}}^{\chi_0}(x)|\tau_1-\tau_2|)}{4|\omega_{\text{eff}}^{\chi_0}(x)|^3}e^{-\omega_{\text{eff}}^{\chi_0}(x)|\tau_1-\tau_2|} \delta(x_1-x_2) \delta(x-x_1)\,. \nonumber
\end{align}
We see that~\eqref{eq:10} agrees with \eqref{eq:8} to first order in perturbation theory.  Thus we have shown that
\begin{align}
  \label{eq:11}
\lim_{\omega_3\to 0}&\left( \lim_{\beta \to \infty}\mathbb{E}_{\chi \sim P_\beta}\!\left[ \omega_3\langle 0^\chi| \widehat{Q}(\omega_3, x_3)\widehat{\phi}(\tau_1, x_1)\widehat{\phi}(\tau_2, x_2)|0^\chi\rangle\right]\right)
\\
&\nonumber \qquad \qquad \qquad \qquad \qquad = \lim_{\beta \to \infty}\mathbb{E}_{\chi \sim P_\beta}\!\left[\frac{\delta}{\delta \chi(x_{3})}\langle 0^\chi|\widehat{\phi}(\tau_1, x_1)\widehat{\phi}(\tau_2, x_2)|0^\chi\rangle\right]\,,
\end{align}
in accordance with the more general equation~\eqref{E:altogether1}.

\subsection{Perturbative electric scalar QED}
\label{S:electricQED}

\subsubsection{Simple correlation functions}

Now we turn to Carrollian scalar QED in $d+1$ spacetime dimensions at finite temperature as described by the imaginary-time action
\begin{align}
\label{E:lag1}
S= a^d\sum_{k \in \Gamma}  \left( |\partial_{\tau}\phi_k + i \,a^{d/2} g A_{\tau,k} \phi|^2 + m^2 |\phi|^2 + \frac{1}{2}(\partial_{\tau}A_k - \nabla A_{\tau,k})^2 +\frac{1}{2\xi}(\nabla \cdot A_k)^2\right)\,.
\end{align}
Here we have regularized the theory by placing it on a spatial hypercubic lattice with lattice spacing $a$; $A_{\tau,k}$ refers to the time-component of the gauge field at site $k$; the vector potential $A_k$ and spatial derivatives $\nabla$ are a shorthand for their discretized versions as discussed in Subsection~\ref{S:pureEM}; and $g$ is a rescaled electromagnetic coupling which we hold finite in the continuum limit. The bare coupling is $g a^{d/2}$, which can be viewed as a multiplicative renormalization of $g$ which renders the theory UV-finite. We also work in an $R_\xi$ version of Coulomb gauge which produces the last term. 

We will take $\xi=0$, i.e.\ Coulomb gauge. Because the coupling between the temporal component $A_{\tau}$ and the vector potential (really the link variables of the discretized theory) goes like the time derivative of the gauge-fixing condition, $A_{\tau}$ and the vector potential decouple in this gauge choice. As a result $A_{\tau}$ and the matter field make up a closed sector, and the vector potential makes another. The latter is completely soluble and was discussed in Subsection~\ref{S:pureEM}. The former has a simple continuum limit in which lattice perturbation theory coincides with a continuum prescription, as for the scalar Yukawa theory and the electric theories studied in Subsection~\ref{S:electric2}. 

In that prescription the tree level $A_{\tau}$ propagator is the $a\to 0$ limit of the lattice propagator for $A_{\tau}$, and reads $\langle A_{\tau}(\tau,x)\,A_{\tau}(\tau',y)\rangle_0 = \frac{\Gamma(\frac{d-2}{2})}{4 \pi^{d/2}}\frac{1}{|x-y|^{d-2}}\,\delta(\tau - \tau')$.  Similar to the $\chi \chi$ two-point function in the Yukawa setting, here there is a divergence at $x = y$ which is regulated by a lattice. We parameterize the short-distance limit as $ \kappa /a^d$ where $\kappa$ is a non-universal coefficient depending on the lattice regularization.  As before, the coefficient $\kappa$ will survive in the $a \to 0$ limit and as such comprises UV data which is required to define the model. 

We begin by computing $\langle \phi^*\phi\rangle$ to one-loop, corresponding to the diagrams
\begin{center}
\includegraphics[scale=.5]{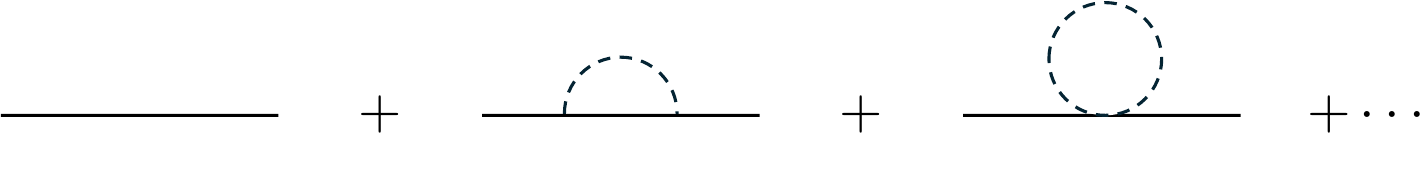}
\end{center}
where solid lines correspond to $ \phi$ propagators and dashed lines correspond to $A_{\tau}$ propagators.  The result in the $a \to 0$ limit is
\begin{align}
\label{E:phiphitot}
&\langle \phi^*(\tau,x) \phi(\tau',y)\rangle = K_\beta(\tau - \tau')\,\delta(x-y) \\
& \qquad \qquad \qquad \qquad \qquad + \frac{g^2 \kappa}{2m} \,\text{coth}\!\left(\frac{\beta m}{2}\right)\left(1 + 2m^2 \frac{\partial}{\partial m^2}\right)\! K_{\beta}(\tau - \tau') \delta(x-y)+\,O(g^4)\,, \nonumber
\end{align}
where $K_{\beta}(\tau)$ was defined in~\eqref{E:Kbeta} and where we note the presence of $\kappa$ in the 1-loop correction.  Above we have assumed that $|\tau - \tau'| \leq \beta$, and we will continue to make this assumption in our expressions below. Note that the loop correction remains ultra-local. We caution that~\eqref{E:phiphitot} is not gauge-invariant since $\phi^*$ and $\phi$ carry non-trivial charge, and as such the operators should be dressed by Wilson lines.  We will return to this consideration in our Hamiltonian analysis below.

Next we turn to $\langle A_{\tau}  A_{\tau}\rangle$, given by 
\begin{center}
\includegraphics[scale=.5]{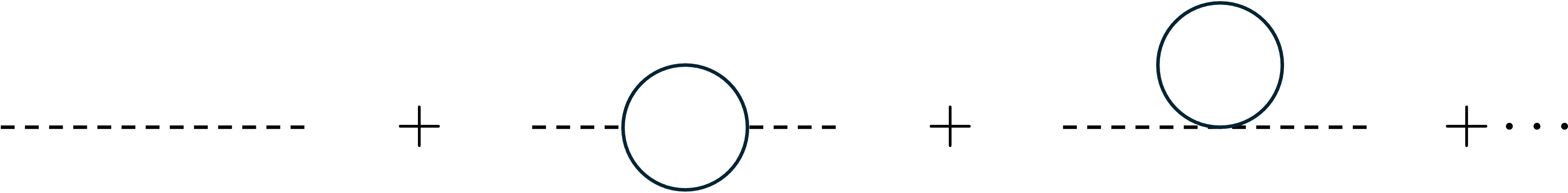}
\end{center}
where we have used the same diagrammatic notation as above. In the $a \to 0$ limit we find
\begin{align}
\label{E:AtAttot}
\langle A_{\tau}(\tau,x) A_{\tau}(\tau',y)\rangle = \frac{\Gamma(\frac{d-2}{2})}{4 \pi^{d/2}}\frac{1}{|x-y|^{d-2}}\,\delta(\tau-\tau') + \frac{g^2}{1 - \cosh(\beta m)} \frac{\Gamma(\frac{d-4}{2})}{16 \pi^{d/2}}\,\frac{1}{|x-y|^{d-4}}\, + O(g^4)\,,
\end{align}
which is devoid of $\kappa$-dependence.  We see that the loop corrections have progressively worse divergences in the $|x-y| \to \infty$ limit, suggesting that the theory should most naturally live on a compact manifold or otherwise have some other form of natural IR cutoff.  We remark that the one-loop correction above, by virtue of Ward identities, can be interpreted as a vacuum polarization coming from the two-point function $\langle J^{\tau} J^{\tau}\rangle$ of the pure matter theory.

In similar spirit we can compute the correlator of the electric field. The leading order contribution, the same as the result in pure Carrollian electromagnetism, vanishes at zero temperature and the corrections are due to interactions between $A_{\tau}$ and the matter and so are easily calculable in the continuum limit. The result is 
\begin{align}
\label{E:EEcorr1}
\langle E_i(\tau,x)\,E_j(\tau',y)\rangle &=\langle E_i(\tau,x) E_j(\tau',y)\rangle_0
\\
&\nonumber \qquad \qquad - (d-4)(d-3)\frac{g^2}{1 - \cosh(\beta m)} \frac{\Gamma(\frac{d-4}{2})}{16 \pi^{d/2}}\,\frac{(x_i - y_i)(x_j - y_j)}{|x-y|^{d-2}}\, + O(g^4) \,,
\end{align}
at one loop order. Note that at zero temperature and nonzero mass $m$ this two-point function vanishes: the free-field correlation function vanishes in the limit, on account of the conservation of the electric field and the mass gap to excitations thereof, and the loop correction dies off exponentially fast too.

A natural object in Carrollian scalar QED is the normally-ordered bilocal
\begin{align}
\mathcal{O}(\tau,x) = a^{d/2}\,:\phi^*(\tau,x)\,\phi(\tau,x):\,.
\end{align}
The bilocal is charge neutral and as such does not need to be dressed by a Wilson line.  The factor of $a^{d/2}$ is required to render its correlations UV-finite. At two loops its expectation value is given by
\begin{center}
\includegraphics[scale=.5]{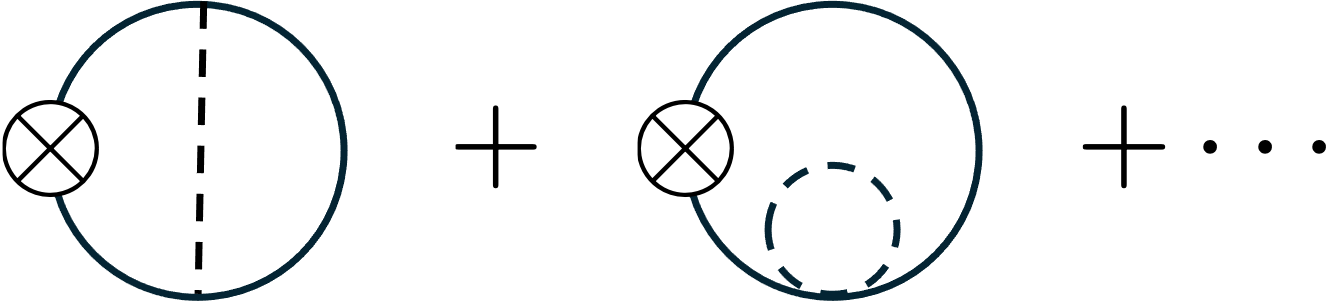}
\end{center}
(where we note that the omitted loop diagram is zero due to normal ordering) and in the $a \to 0$ limit we obtain
\begin{align} \label{E:Oexpt}
\langle \mathcal{O} \rangle =  a^{-d/2}\,\frac{1}{2m} \coth\!\left(\frac{\beta m}{2}\right) \left(1 + \frac{g^2 \kappa}{2}\,\frac{\beta}{2}\left[\tanh\!\left(\frac{\beta m}{2}\right) - \coth\!\left(\frac{\beta m}{2}\right)\right]+\,O(g^4)\right)\,,
\end{align}
where the factor of $a^{-d/2}$ arises from the lattice identity $\delta(0)=a^{-d}$. This comports with our results for pure electric theories and the scalar Yukawa theory, where one-point functions of normalized operators diverge as $a^{-d/2}$ in the continuum limit. This divergence is not as bad as in the Yukawa setting, since it can naturally removed by a slightly different normal-ordering prescription, as we explore in the Hamiltonian formalism below.  For the connected two-point function of $\mathcal{O}$ at two-loop order we find
\begin{center}
\includegraphics[scale=.5]{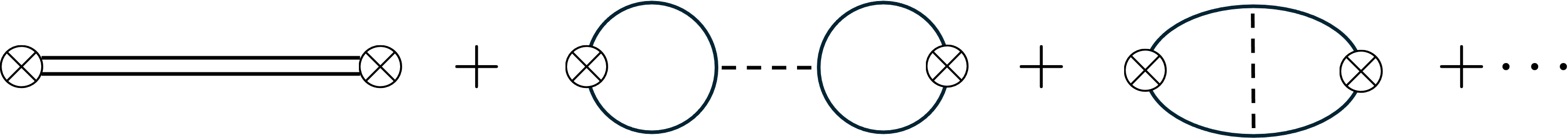}
\end{center}
which in the $a \to 0$ limit is finite and reads
\begin{align}\label{E:OOcorr1}
\langle \mathcal{O}(\tau,x) \mathcal{O}(\tau',y)\rangle_{\text{conn}} = K_\beta(\tau - \tau')^2\,\delta^d(x-y)  \,+ O(g^4)\,.
\end{align}
There is a delicate cancellation at loop order which leaves us with the tree level result.

Our above diagrammatics show us that the 1-loop $\beta$-functions of $g$ and $m$ are trivial, i.e.~that they maintain their classical scalings.  This accords with our expectations about Carrollian scalar QED being akin to a 0+1 quantum mechanical theory.

As in the case of the Yukawa theory, we can also consider the $\beta\to\infty$ limit of these results. In this limit we find
\begin{align}
  \begin{split}\label{eq:QEDzeroT}
    \langle \phi(\tau,x)\phi(\tau',y)\rangle&=\frac{e^{-m|\tau_{12}|}}{2m}\delta(x-y)\left(1-\frac{g^2\kappa}{2}|\tau_{12}|\right)+O(g^4)\,,\\
    \langle A_\tau(\tau,x)A_\tau(\tau',y)\rangle&=\frac{\Gamma(\frac{d-2}{2})}{4 \pi^{d/2}}\frac{1}{|x-y|^{d-2}}\,\delta(\tau-\tau')+O(g^4)\,,\\
    \langle E_i(\tau,x)E_j(\tau',y)\rangle&=0+O(g^4)\,,\\
    \langle \mathcal{O}\rangle&=a^{-d/2}\frac{1}{2m}+O(g^4)\,,\\
    \langle \mathcal{O}(\tau,x) \mathcal{O}(\tau',y)\rangle_{\rm conn}&=\frac{1}{4m^2}e^{-m|\tau-\tau'|}+O(g^4) \,.
 \end{split}
\end{align}
It is easy to verify that connected $n$-point functions with $n>2$ vanish in the continuum limit. In fact they have the same scalings as connected $n$-point functions in pure electric theories, going as $a^{\frac{d(n-2)}{2}}$. This fact is manifest for tree-level processes, but our scaling limit ensures that this scaling is preserved at loop level.

\subsubsection{Hamiltonian formalism}

Here we pursue a Hamiltonian formulation the Carrollian limit of scalar QED, which clarifies the structure of our perturbative path integral calculations above.  The Hamiltonian is
\begin{align}
H = \sum_{i\,\in\,\Gamma}\left(\frac{|\pi_i|^2}{a^d} +a^d\left(  m^2 |\phi_i|^2 + \frac{1}{2} \sum_{m=1}^dE_{i,m}^2\right)\right),
\end{align}
where $\pi$ is the canonical conjugate of $\phi$ obeying
\beq
	[\phi_i,\pi_j] = i \,\delta_{ij}\,.
\eeq
The Hamiltonian is subject to the constraint imposed by the Gauss' Law 
\begin{align}
\nabla \cdot E = i g a^{d/2} (\pi \phi^* - \pi^* \phi) \,,
\end{align}
which is implemented in e.g.~the phase space path integral by a Lagrange multiplier corresponding to $A_\tau$. Here $\nabla\cdot E$ refers to the discretized version of the divergence. The Hamiltonian form of the theory makes manifest that $\phi$ and $E_j$ are free fields. To write down states in the theory it is simplest to work with the oscillator representation $\phi(x) = \frac{a^{-d/2}}{\sqrt{2m}} \left(a_1^\dagger(x) + a_2(x) \right)$ and $\pi(x) = -ia^{d/2}\sqrt{\frac{m}{2}} \left(a_1(x) - a_2^\dagger(x) \right)$, so that the local $U(1)$ charge is 
\begin{equation}
    Q(x) = i \left(\pi(x) \phi^*(x) - \pi^*(x) \phi(x)\right)
    = a_2^\dagger(x) a_2(x) - a_1^\dagger(x) a_1(x)
    = N_2(x) - N_1(x)\,.
\end{equation}
Here $N_i(x) = a^\dagger_i(x) a_i(x)$ is the occupancy number of the $a_i$--type oscillators, and we readily see that $\phi$ has charge $-1$. Since the Hamiltonian commutes with the electric field, the energy eigenstates are also eigenstates of $\vec{E}$. We can then write down the energy eigenstates as 
\begin{equation}
    \ket{N_1(x),N_2(x),E(x)}\,,
\end{equation}
where the electric field satisfies the constraint
\begin{equation}
\nabla \cdot E = g a^{d/2} \,Q(x) = g a^{d/2} \,(N_2(x) - N_1(x))\,.
\end{equation}
This constraint implies that $E$ can be separated into a unique curl-free part, and the curl of some vector field. The curl-free part is the classical electrostatic solution in the presence of a charge distribution $Q(x)$, while the curl part is an additional tower of excited states built on top of the electrostatic solution, in exact correspondence to the states built on the empty vacuum in Subsection \ref{S:pureEM}.

The energy of each state $\ket{N_1(x),N_2(x),E(x)}$, subtracting the zero point vacuum energy, is
\begin{equation}
    \mathcal{E}(N_1(x), N_2(x), E(x)) = \sum_{i\in \Gamma} \left(\frac{a^{d}}{2}\sum_{m=1}^dE_{i,d}^2 + m N_1(x_i) + m N_2(x_i)\right).
\end{equation}
 The ground state is $\ket{0,0,0}$, and there exists a gap of order $m$ to the first excited states of the matter and of order $(ga)^2$ for the electric field.

For simplicity, we will compute correlation function in the Hamiltonian picture at zero temperature so that we can project the in-- and out--states to the vacuum.  To begin, we observe that the vacuum correlation functions of $\vec{E}$ are zero, in agreement with the zero temperature limit of \eqref{E:EEcorr1} up to a contact term.  Now for the operator 
\begin{equation}
    \mathcal{O}(x) = a^{d/2} : \phi^*(x) \phi(x):\, = \frac{a^{-d/2}}{2m}:\left( a_1^\dagger(x) a_2^\dagger(x) + a_1(x) a_2(x) + a_1^\dagger(x) a_1(x) + a_2(x) a_2^\dagger(x)  \right):\,,
\end{equation}
the expectation value $\left<\mathcal{O} \right>$ depends on the definition of the normal ordering constant. One typically defines the normal-ordered operator as moving the creation operators to the left, in which case $\left<\mathcal{O} \right> = 0$. However without doing so we obtain
\begin{equation}
    \left<a^{d/2} \phi^*(x) \phi(x) \right> = \frac{a^{-d/2}}{2m}\left< a_1^\dagger(x) a_2^\dagger(x) + a_1(x) a_2(x) + a_1^\dagger(x) a_1(x) + a_2(x) a_2^\dagger(x)  \right>
    =\frac{1}{2ma^{d/2}}\,,
\end{equation}
diverging as $a^{-d/2}$ but in agreement with the zero temperature limit of \eqref{E:Oexpt}.

Turning to the two-point functions of $\mathcal{O}$, we find
\begin{equation}
\langle \mathcal{O}(\tau,x) \mathcal{O}(0,y) \rangle = \frac{a^{-d}}{4m^2} \langle 1_x,1_x,0| e^{-H\tau} | 1_y,1_y,0 \rangle = \frac{e^{-2 m |\tau|}}{4m^2}\,\delta(x-y)\,,
\end{equation}
in agreement with \eqref{E:OOcorr1}.  The Hamiltonian formalism elucidates the cancellations we observed in our diagrammatics in the path integral formulation: since $\mathcal{O}$ carries no charge and so does not alter electric field, its tree level two-point function is exact even at finite temperature and couplings.

Finally, we can examine the two-point function of $\phi^*$ and $\phi$. We recall that $\phi^*$ and $\phi$ alone are not gauge-invariant operators, while $\mathcal{O}$ and $E$ are gauge-invariant. To make e.g.~$\phi$ gauge invariant we must dress it with a Wilson line.  For example the operator $\Phi(\tau,x) = \phi(\tau,x)\,e^{ia^{d/2}g \int_{\tau}^{\infty} d\tau' ~A_{\tau}(\tau',x)}$ is gauge-invariant, with the temporal Wilson line enforcing the Gauss's law constraint in the presence of the charge. We also see that 
\begin{equation}
    \Phi(0,x) \ket{0,0,0} = \frac{a^{-d/2}}{\sqrt{2m}}\ket{0,1_x,E_x},
\end{equation}
where the electric field satisfies 
\begin{equation}
    \nabla \cdot \vec{E}= - ga^{d/2} \,\delta^d(x-y)\,.
\end{equation}
In the continuum flat space limit the (curl free) solution to this equation is
\begin{equation}
    E^j(y) = ga^{d/2}\frac{\Gamma(d/2)}{2\pi^{d/2}}\,\frac{x^j - y^j}{|x-y|^{d}}.
\end{equation}
and so we can compute the 2-point function of $\Phi$ as
\begin{align}
\begin{split}
\langle \Phi(\tau,x) \Phi(0,y) \rangle &= \frac{a^{-d}}{2m} \langle0,1_x,E_x| e^{-H\tau}|0,1_y,E_y \rangle  \\
    &=  \frac{e^{-m|\tau|} }{2m}  e^{-  \frac{g^2 \Gamma(\frac{d-2}{2})}{8 \pi^{d/2}} a^{2} |\tau|} \delta(x-y) \\
    &= \frac{ e^{-m |\tau|} }{2m}e^{- \frac{g^2 \kappa }{2} |\tau| }\delta(x-y)\,.
\end{split}
\end{align}
Expanding this out to order $g^2$, we can compare with the $\beta \to \infty$ limit of~\eqref{E:phiphitot} given in \eqref{eq:QEDzeroT} and find a match.

Overall we see that the computations in the Hamiltonian framework, at least for $\beta \to \infty$, are very simple, enabling us to compute correlators directly at all orders in the coupling.

\section{Towards Carrollian holography}
\label{sec:towards-carr-hologr}

In this paper we have explored quantum-mechanical properties of Carrollian theories, with an eye towards their application to flat space holography.  We conclude by outlining some structural connections between Carrollian theories and flat space quantum gravity, and highlight opportunities for further inquiry in the form of two important challenges for Carrollian holography. 

\subsection{Matching symmetries}
\label{sec:electr-magn-sect}

A natural place to begin discussing a candidate realization of flat space holography is symmetry. In the AdS/CFT correspondence bulk isometries act as global symmetries on the conformal boundary, with the isometries of AdS$_{d+1}$ space generating the action of the conformal group $SO(d,2)$ at infinity. Flat space quantum field theory in $d+2$ spacetime dimensions is invariant under the Poincar\'e group $ISO(d+1,1)$. This symmetry is enhanced in flat space quantum gravity, and the precise form of the ensuing asymptotic symmetry group is a matter of current research. In 3+1 bulk dimensions this group includes at least the Bondi-van der Burg-Metzner-Sachs (BMS) group, generated by Lorentz transformations and supertranslations, which are labeled by a function on the sphere at celestial infinity~\cite{Bondi:1962px,  Sachs:1962zza}. There is reason to think that the correct symmetry group is enhanced further to the extended BMS group, generated by the BMS group together with superrotations~\cite{Barnich:2009se,Barnich:2010eb}, or arbitrary diffeomorphisms of the asymptotic sphere~\cite{Campiglia:2014yka,Campiglia:2015yka}, or perhaps even $w_{1+\infty}$~\cite{Strominger:2021mtt,Ball:2021tmb} (which has an interesting interplay with twistor space~\cite{Adamo:2021lrv} and Carroll symmetries~\cite{Saha:2023abr,Mason:2023mti,Donnay:2024qwq}).

The crux of the Carrollian proposal for flat space holography is that the dual to flat space gravity is a Carrollian theory living at infinity. The best developed dictionary at the moment relates the scattering amplitudes of massless particles to correlation functions of a single Carrollian theory living on future (or past) null infinity. We will have more to say later in this Discussion about the single copy of null infinity (as opposed to all of null infinity), and the ignorance about how massive particles fit into the story.

How does the Poincar\'e group, and more generally the BMS group and its extensions, act on null infinity? The Poincar\'e group acts as the conformal extension of the Carroll group. Supertranslations act as a position-dependent time translation $u\to u+f(x)$ (what we have called supertranslations in the body of this manuscript), and superrotations act as local conformal transformations. Moreover, the soft graviton and subleading soft graviton theorems indicate that the supertranslation and superrotation symmetries are spontaneously broken in $3+1$ bulk dimensions. There is a manifold of soft vacua for the graviton generated by the extended BMS group modulo its Poincar\'e subgroup~\cite{Ashtekar:1981hw,He:2014laa,Kapec:2014opa}. Similar statements hold in 2+1 bulk dimensions~\cite{Barnich:2014kra,Barnich:2015uva}.

Flat space quantum gravity exhibits a tower of classical soft theorems beyond supertranslations and superrotations. These correspond to asymptotic symmetries that grow with positive powers of the radial coordinate and whose action on null infinity is thus ill-defined.

With this in mind the natural expectation is that a Carrollian dual to flat space gravity is a conformal Carrollian field theory (sometimes called a CCFT, which confusingly is also the abbreviation used for celestial CFT), with suitably spontaneously broken supertranslation and superrotation symmetries. Comparing and contrasting with the result of this manuscript in quantizing Carrollian field theories -- that our theories lack a superrotation symmetry, have unbroken supertranslations, and typically have a mass scale -- we land on a first challenge for any realization of putative Carrollian holography: \newline

\noindent \textbf{Challenge 1:} Identify a family of conformal Carrollian CFTs with not only supertranslation symmetry, but superrotation invariance. Furthermore these extended symmetries ought to be spontaneously broken to the conformal Carroll subgroup.

\subsection{From scattering amplitudes to Carrollian correlators}

Should there be a realization of Carrollian holography there is a well-motivated dictionary~\cite{Donnay:2022aba,Donnay:2022wvx,Bagchi:2022emh,Kim:2023qbl,Jain:2023fxc} relating the scattering amplitudes of massless particles to correlation functions of a dual CCFT. This dictionary has been landed on from several points of view, including the action of Poincar\'e symmetry at null infinity~\cite{Donnay:2022aba,Donnay:2022wvx} and adapting the GKPW construction of AdS/CFT~\cite{Kim:2023qbl,Jain:2023fxc}. 

Let us give a brief summary of the approach taken in~\cite{Donnay:2022aba,Donnay:2022wvx} in $3+1$ dimensions.  Consider the scattering amplitudes of a massless field $\Phi$. Expanding it in plane wave creation/annihilation operators $a_k^{\dagger}$ and $a_k$ close to future null infinity
\begin{equation}
   \Phi(x)=\int\frac{d^3 k}{(2\pi)^3 2k^0}\left(a_ke^{ik\cdot x}+a^\dagger_ke^{-ik\cdot x}\right)\,,
\end{equation}
a saddle point approximation yields the result at large $r$~\cite{He:2014laa,Bekaert:2022ipg,Donnay:2022wvx}
\begin{equation}
\label{eq:saddlepoint}
    r\Phi(x)\approx \Phi_0(u,x^i)= \int^{\infty}_0 d\omega \left(e^{-i\omega u}a_{\omega q(z,\bar{z})}-e^{i\omega u}a^\dagger_{\omega q(z,\bar{z})}\right),
\end{equation}
where $q^\mu$ describes a unit vector pointing to an angle on the celestial sphere. Under this relation we can trade the four-momentum of an outgoing or ingoing massless particle for a frequency $\omega$ and an angle on the sphere. It should be noted that this saddle-point approximation is valid only under the assumption $\omega>0$, so that \eqref{eq:saddlepoint} only applies for hard modes.

The field $\Phi_0$ and its higher-helicity $s$ analogues, which we notate as $\phi_{\Delta,s}$, transform as conformal Carrollian primaries with conformal weight $\Delta=1$ under conformal Carrollian symmetries \cite{Donnay:2022aba,Bagchi:2022emh, Chen:2021xkw}. These act on future null infinity via vector fields
\begin{equation}
    \xi= \left(\mathcal{T}(z,\bar{z})+\frac{u}{2}(\partial Y(z)+\bar{\partial}\bar{Y}(\bar{z}))\right)\partial_u+Y(z)\partial_z+\bar Y(\bar z)\partial_{\bar{z}} \, ,
\end{equation}
where we are using flat complex coordinates $z$ on the celestial sphere, $\mathcal{T}$ is a supertranslation and $Y$ a superrotation; on past null infinity they act in the same way up to an antipodal flip. On $\Phi_0$ and its higher-helicity analogues they act as
\begin{align}
  \label{eq:2}
  \delta_{\mathcal{T},Y,\bar{Y}}\phi^\varepsilon_{\Delta,s}= \left[  \left(\mathcal{T}+\frac{u}{2}(\partial Y+\bar{\partial}\bar{Y})\right)\partial_u+Y\partial  +\tfrac{\Delta+\varepsilon s}{2}\partial Y+ \bar Y\bar\partial  +\tfrac{\Delta-\varepsilon s}{2}\bar\partial\bar Y  \right]   \phi^\varepsilon_{\Delta,s}  \,,
\end{align}
where the label $\varepsilon=\pm 1$ distinguishes between out/ingoing asymptotic fields. A curious feature of the representation theory of the conformal Carroll group is that, if $\phi_{\Delta,s}$ is a primary, then so is $\partial_u \phi_{\Delta,s}$ but now with dimension $\Delta+1$.\footnote{Note that the representation \eqref{eq:2} restricted to the Poincaré subgroup is nothing but the standard Wigner representation for massless particles after applying a Fourier transform~\cite{Bekaert:2022ipg,Nguyen:2023vfz}.}

The proposed dictionary \cite{Donnay:2022aba,Donnay:2022wvx} between scattering amplitudes and conformal Carroll correlators $\mathcal{C}_{n}$ amounts to the combination of Fourier transforming with respect to frequency and applying an antipodal map to the angle of ingoing particles:
\begin{align}
  \langle\phi^{\varepsilon_1}_{\Delta=1}(u_1,z_1,\bar{z}_1)\cdots \phi^{\varepsilon_n}_{\Delta=1}(u_n,z_n,\bar{z}_n)\rangle
  &=\prod^{n}_{i=1}
  \left(
  \int^\infty_0\frac{d \omega_i}{2\pi} e^{i\varepsilon_i\omega_i u_i}
  \right)
  \,\mathcal{A}(\omega_1,z_1,\bar{z}_1, \varepsilon_1; \cdots ;\omega_n,z_n,\bar{z}_n, \varepsilon_n) \nonumber\\
  & =\mathcal{C}_{n}(u_1,z_{1},\bar{z}_1,\varepsilon_{1};\cdots; u_n,z_n,\bar{z}_n,\varepsilon_{n})\,,
  \label{eq:mom-to-carr}
\end{align}
where we hide the dependence on the helicities and it is understood that we act with the antipodal map on the positions $(z,\bar{z})$ of incoming particles. This transform translates scattering amplitudes from the momentum basis into conformal Carrollian correlators. It can be checked that the resulting object transforms as a correlator of primary operators \eqref{eq:2} of weight $\Delta=1$.\footnote{One can modify the dictionary presented in~\eqref{eq:mom-to-carr} so that the operator insertions carry arbitrary dimension $\Delta$ by instead performing  the modified Mellin transform~\cite{Banerjee:2018gce,Bagchi:2022emh} $\int^\infty_0\!\frac{d\omega}{2\pi}\, \omega^{\Delta-1}  e^{i\varepsilon \omega u}$ on each external state.} For instance, applying the map to a two-point scattering amplitude of scalars in 3+1 dimensions
\begin{equation}
  \label{eq:2pointmom}
  \mathcal{A}_{2}(\omega_1,z_1,\bar{z}_1, +;\omega_2,z_2,\bar{z}_2,-)=\pi\frac{\delta(\omega_1-\omega_2)}{\omega_1}\delta(z_{12})\delta(\bar{z}_{12})\,,
\end{equation}
we obtain
\begin{equation}
  \label{eq:2pt-carr}
  \mathcal{C}_2(u_{1},z_1,\bar{z}_1,+;u_2,z_2,\bar{z}_2,-)=\tfrac{1}{4\pi}\delta(z_{12})\delta(\bar{z}_{12})\int^\infty_0\frac{d \omega}{\omega}\,e^{i\omega(u_1-u_2)}
\end{equation}
where we are using the notation $z_{12}=z_{1}-z_{2}$. This expression is infrared-divergent due to the pole at $\omega=0$. Nevertheless, after differentiating with respect to $u_1$ and $u_2$ to project out this zero-mode -- for which the transformation \eqref{eq:saddlepoint} was not valid -- one arrives at 
\begin{equation}
\pd_{u_{1}}\pd_{u_{2}}\mathcal{C}_2(u_{1},z_1,\bar{z}_1,+;u_2,z_2,\bar{z}_2,-)= \tfrac{1}{4\pi}\delta(z_{12})\delta(\bar{z}_{12})\lim_{\epsilon\to 0^{+}}\frac{1}{(u_{21}-i\epsilon)^2}\,,
\end{equation}
which is ultralocal on the sphere. So two-point amplitudes correspond to purely electric correlators. Similarly, three-point amplitudes in $\Phi^3$ theory map to purely electric three-point functions \cite{Bagchi:2023fbj,Nguyen:2023miw} under this dictionary,\footnote{This is the three-point function in Lorentzian signature that corresponds to collinear scattering in the bulk. Other three-point functions can be obtained by analytic continuation in the spatial coordinates as in \cite{Salzer:2023jqv,Mason:2023mti}.} namely
\begin{equation}
  \label{eq:6}
  \mathcal{C}_3= g u^a_{12}u^{b}_{23}u^{c}_{31}\delta(z_{12})\delta(z_{23})\delta(\bar{z}_{12})\delta(\bar{z}_{23}),\qquad a+b+c=4-\sum_{i}\Delta_i\,,
  \end{equation}
  where $g$ is the cubic coupling. For further explicit expressions including for example four-point functions see \cite{Bagchi:2022emh,Donnay:2022aba,Donnay:2022wvx,Bagchi:2023fbj,Salzer:2023jqv,Bagchi:2023cen,Nguyen:2023miw,Mason:2023mti,Banerjee:2024hvb}. Crucially, only two- and three-point functions are ultralocal. Reconstructing $n$-point functions with $n>3$ using the dictionary leads to correlations with non-trivial spatial dependence. In fact, the correlators so obtained are not supertranslation-invariant.
  
While the dictionary~\eqref{eq:mom-to-carr} will generate non-trivial $n$-point functions for all $n$ in a putative dual CCFT, the Carrollian theories studied in this manuscript have only Gaussian correlations. This leads to the next challenge motivated by our work: \newline

\noindent \textbf{Challenge 2:} Identify CCFTs with tractable non-Gaussianity. \newline

Before going on, we would like to comment briefly on the question of whether a dual should exist along all of null infinity, or only along its future (or past) half. The dictionary~\eqref{eq:mom-to-carr} suggests the dual only lives along the future half. While this is at odds with the intuition developed from AdS/CFT, it follows from the flat space analogue~\cite{Kim:2023qbl,Jain:2023fxc} of the GKPW algorithm. The important fact is that in flat space physics we impose quite different boundary conditions than in a space with a timelike boundary like AdS. In particular, we impose that particles are incoming in the past and outgoing in the future. From the point of view of identifying sources and conjugate operators from the near-boundary data of a massless field, this amounts to a non-local split whereby we have two sets of positive-frequency operators, one coming from past null infinity and the other from the future (since the energy of asymptotic states is always bounded below). On time-reversing the modes along either past or future infinity, we can group them together to restore a local-in-null-time description precisely through the dictionary~\eqref{eq:mom-to-carr}. This demonstrates that we only have a local-in-null description along either past or future null infinity.

\subsection{Soft modes and magnetic correlators}

In addition to time-dependent solutions needed to reproduce scattering amplitudes via the dictionary~\eqref{eq:mom-to-carr}, the conformal Carrollian Ward identities also allow for time-independent solutions. As is clear from the expression~\eqref{eq:2}, the resulting correlators are in form identical to those of a $d$-dimensional Euclidean CFT.  This is clearly reminiscent of our discussion of the $\chi$ correlators of the magnetic Carrollian theories. 

The dictionary~\eqref{eq:mom-to-carr} relates soft insertions in scattering amplitudes to magnetic operators in a putative Carrollian dual. The correlators with magnetic insertions so obtained are typically afflicted with IR divergences. 

As an example consider flat space gravity. As discussed above, the asymptotic symmetries of an asymptotically flat spacetime include the BMS group.  When studying perturbative scattering amplitudes the choice of a vacuum spontaneously breaks the BMS group down to its Poincar\'e subgroup \cite{Ashtekar:1981hw}. Consequently, one finds that the gravitational vacuum is labelled by a Goldstone boson-like operator $C(z,\bar{z})$ that shifts under supertranslations; see, e.g.~\cite{Kapec:2022hih} for a discussion in the context of the celestial CFT proposal. The two-point correlator of these Goldstone bosons in $3+1$ bulk dimensions is
\begin{equation}
    \langle C(z_1,\bar z_1)C(z_2,\bar z_2)\rangle=\frac{4G_N}{\pi}\log \!\left(\tfrac{\Lambda_{\rm IR}}{\mu} \right) |z_{12}|^2  \log |z_{12}|^2,
\end{equation}
where $\Lambda_{\rm IR}$ is an IR cut-off and $\mu$ is a fiducial scale separating hard from soft modes. This correlator can be determined either from matching to Weinberg's factorization formula of gravitational IR divergences \cite{Himwich:2020rro,Kapec:2021eug,Kalyanapuram:2021bvf} or from an examination of boundary contributions to the variational principle \cite{Nguyen:2021ydb}. Similar results hold in QED \cite{Arkani-Hamed:2020gyp,He:2024ddb} and QCD \cite{Magnea:2021fvy,Gonzalez:2021dxw} (to leading order in $\alpha_s$) where one finds analogous infinite-dimensional asymptotic symmetry groups that are spontaneously broken. For superrotations an explicit (magnetic) Carrollian action for the corresponding Goldstone has been constructed in \cite{Nguyen:2020hot}.

The canonically conjugate operator to $C$ is the soft graviton $N$ \cite{He:2014laa} (with analogous results holding again for QED and QCD). Its action on a given vacuum labelled by $C$ produces a shift in the vacuum $C+\delta C$, so that the soft graviton theorem can be understood as an infinitesimal parallel transport in the space of gravitational vacua
\begin{equation}
  \label{eq:3}
  \langle N(x)\mathcal{O}_1\ldots \mathcal{O}_n\rangle_{C=0}\sim \frac{\delta}{\delta C(x)} \langle\mathcal{O}_1\ldots \mathcal{O}_n\rangle\big|_{C=0}\,.
\end{equation}
See \cite{Kapec:2022hih} for a precise statement.

These properties of the gravitational (or gauge-theoretical) soft sector clearly mirror the properties of the magnetic sector discussed in Section \ref{S:electromagnetic}. As we saw in the case of scalar Yukawa theory the dynamics of the magnetic mode $\chi$ described the sector of vacua thus playing an analogous role to the gravitational Goldstone boson $C$. On the other hand, the action of its canonically conjugate mode $Q$ shifts the vacuum and so is an analogue of the soft graviton operator $N$.

Another indication that soft modes and magnetic Carrollian theories are related can be seen in pure three-dimensional gravity.  Like its AdS version~\cite{Cotler:2018zff}, this model can be rewritten in terms of an effective theory for soft excitations at null infinity.  The action of these modes is the geometric action of the BMS$_3$ group \cite{Barnich:2013yka,Barnich:2017jgw,Merbis:2019wgk} and it is of the magnetic Carrollian type. More explicitly, boundary excitations are labelled by a pair of fields $(\alpha,f)$ with an action~\cite{3dflat}
\begin{equation}
  \label{eq:1}
  S=-\frac{1}{8\pi G}\int d u\, d\varphi \left(\alpha \partial_uP+P\right),\qquad P=\{\tan(f/2),\varphi\}\,,
\end{equation}
where $\{f(\varphi),\varphi\} = \frac{f'''(\varphi)}{f'(\varphi)} - \frac{3}{2}\left( \frac{f''(\varphi)}{f'(\varphi)}\right)^2$ is the Schwarzian derivative of $f$ with respect to $\varphi$. This action is of the magnetic type discussed in Section~\ref{S:magnetic}, although this model has not only a conformal Carroll symmetry, but invariance under supertranslations and superrotations as well. Further details about this theory and its relation to the present manuscript will appear in~\cite{3dflat}.

Taken together, we have provided some evidence of the interpretation
of the sectors in a putative theory of flat space holography. Yet to
make these connections precise it would be necessary to study them in
more detailed in a suitable (toy) model of flat space holography.

\subsection{Massive fields at late times}
\label{sec:carr-theor-as}

This discussion has so far mostly focused on the scattering of massless particles for which there is a proposed dictionary~\eqref{eq:mom-to-carr} with the correlation functions of a putative dual Carrollian CFT at null infinity. How do massive particles fit into this proposal? The answer is currently unknown. We have nothing to contribute on this point, other than to emphasize the importance of this question for practitioners of the field.

We do however wrap up by noting a way that Carrollian physics is related to that of massive fields in flat space, albeit in a way that is quite divorced from the rest of this manuscript. Namely, the dynamics of massive fields near timelike infinity reduces to that of a Carrollian theory on hyperbolic space~\cite{Have:2024dff}. For example, in coordinates where Minkowski space has the line element
\beq
	ds^2 = -d\tau^2 + \tau^2 h_{ij}dx^i dx^j\,,
\eeq
with $h_{ij}$ the metric on a unit-radius $d$-dimensional hyperboloid, the action of a massive scalar $\Phi$ reduces to
\beq
	S[\Phi] = \frac{1}{2}\int d\tau d^{d}x \sqrt{h} \left( (\partial_{\tau} \Phi)^2 - m^2 \Phi^2\right)\,,
\eeq
in the $\tau\to \infty$ limit. This is an ``electric theory'' except on a hyperboloid. This space is sometimes called AdS Carroll~\cite{Figueroa-OFarrill:2018ilb} or Ti~\cite{Figueroa-OFarrill:2021sxz}.\footnote{In a general asymptotically flat spacetimes one expects similarly to find curved versions of Ti at asymptotically late times \cite{Borthwick:2023lye,Borthwick:2024wfn}.}

For a slightly less trivial example consider massive scalar QED in Minkowski space. The Hamiltonian density takes the form
\begin{align}
\begin{split}
  \label{eq:scalarQED}
  \frac{\mathcal{H}}{\sqrt{h}} =\,&|\pi|^2+m^2|\phi|^2+\frac{d\tau^{-1}}{2}\left(\pi^*\phi+\pi\phi^*\right) 
  \\
  &+\tau^{-2}h^{ij}|D_i\phi+ieA_i\phi|^2+\pi^i\pi^jh_{ij}\tau^{2-d}+\frac{1}{4}\tau^{d-4}F_{ij}F_{kl}h^{ik}h^{jl} \, , 
  \end{split}
\end{align}
where we performed the (time-dependent) canonical transformation $\phi\rightarrow \tau^{-d/2}\phi,\pi\rightarrow \tau^{d/2}\pi$ on the scalar field. We can now study the ensuing asymptotic Hamiltonian at late times $\tau\rightarrow \infty$. In particular, for $d=2$ one finds the theory discussed in Section \ref{S:electromagnetic} albeit on hyperbolic space. It would be interesting to see if the tools developed in this work can also be applied to these models.

\subsection*{Acknowledgements}

It is a pleasure to thank Prateksh Dhivakar, Laura Donnay, Yannick Herfray, Andreas Karch, Per Kraus, Alfredo Pérez, Romain Ruzziconi, Aditya Sharma, Wei Song, Andrew Strominger, Alejandro Vilar-López, and Larry Yaffe for valuable discussions. JC is supported by the Simons Collaboration on Celestial Holography. KJ is supported in part by an NSERC Discovery Grant. AR is supported by the U.S. Department of Energy under Grant No. DE-SC0022021 and a grant from the Simons Foundation (Grant 651678, AK). JS is supported by a postdoctoral research
fellowship of the F.R.S.-FNRS (Belgium).

Part of this project was performed during the Programme ``Carrollian
Physics and Holography'' at the Erwin-Schrödinger International Institute
for Mathematics and Physics in April 2024.

\bibliography{refs}
\bibliographystyle{utphys}

\end{document}